\documentclass[manuscript,nonacm]{acmart}
\AtBeginDocument{%
  }

\usepackage{subcaption}

\setcopyright{acmlicensed}
\copyrightyear{2018}
\acmYear{2018}
\acmDOI{XXXXXXX.XXXXXXX}
\acmConference[Conference acronym 'XX]{Make sure to enter the correct
  conference title from your rights confirmation email}{June 03--05,
  2018}{Woodstock, NY}
\acmISBN{978-1-4503-XXXX-X/2018/06}

\begin{document}

\title[AI Adoption for Cross-Cultural Adaptation]{Understanding How International Students in the U.S. Are Using Conversational AI to Support Cross-Cultural Adaptation}

\author{Laleh Nourian}
\orcid{0000-0002-5214-794X}
\affiliation{%
  \institution{Rochester Institute of Technology}
  \city{Rochester}
  \state{NY}
  \country{USA}}
\email{ln2293@rit.edu}

\author{Anisa Callis}
\orcid{0009-0007-3322-3285}
\affiliation{%
  \institution{Rochester Institute of Technology}
  \city{Rochester}
  \state{NY}
  \country{USA}}
\email{amc1672@rit.edu}

\author{Stephanie Patterson}
\orcid{0009-0003-3387-2454}
\affiliation{%
  \institution{Rochester Institute of Technology}
  \city{Rochester}
  \state{NY}
  \country{USA}}
\email{slp9101@rit.edu}

\author{Jadeline Miao}
\orcid{0009-0005-4983-6014}
\affiliation{%
  \institution{Rochester Institute of Technology}
  \city{Rochester}
  \state{NY}
  \country{USA}}
\email{jm2386@rit.edu}

\author{Jamison Heard}
\orcid{0000-0001-6860-0844}
\affiliation{%
  \institution{Rochester Institute of Technology}
  \city{Rochester}
  \state{NY}
  \country{USA}}
\email{jrheee@rit.edu}

\author{Garreth W. Tigwell}
\orcid{0000-0002-3123-6361}
\affiliation{%
  \department{School of Information}
   \institution{Rochester Institute of Technology}
  \city{Rochester}
  \state{NY}
  \country{USA}}
\email{garreth.w.tigwell@rit.edu}

\renewcommand{\shortauthors}{Nourian et al.}

\begin{abstract}
  Moving to a new culture and adapting to a new life, as an international student, can be a stressful experience. In the US, international students face unique overlapping challenges, yet the current support ecosystem, including university support systems and informal social networks, remains largely fragmented. While conversational AI has emerged as a tool used by many (e.g., generative AI chatbots like ChatGPT and Google Gemini), we do not have a clear understanding of how international students adopt and perceive these technologies as support tools. We conducted a survey study (n=60) to map the relationship between international students' challenges and AI adoption patterns, followed by an interview study with 14 participants to identify the underlying motivations and boundaries of use. Our findings show that AI is perceived as a first-aid tool for immediate challenges, however, there is an interest in  transforming AI from a tool for short-term help into a long-term support companion. By identifying where and how AI can provide long-term support, and where it is insufficient, we contribute recommendations for creating AI-powered support tailored to the unique needs of international students.
\end{abstract}

\maketitle


\section{Introduction}

In an era of increasing globalization, moving internationally offers significant personal and professional prospects; however, transitioning to a foreign country as a student can also be an overwhelming experience~\cite{ma2025exploring}. 
The scale of this movement is large; by the end of 2025, the United States had attracted nearly 1.2 million international students,\footnote{Released by Open Doors in November 2025: \url{https://opendoorsdata.org/annual-release/international-students/}} from more than 200 places of origin. For this massive population, the opportunities are often intertwined with various challenges when adapting to life in a new culture, including socio-cultural mismatches, language and communication barriers, academic pressure, daily life and logistical coordination, and interpersonal or psychological difficulties~\cite{lee2025into,ma2025exploring,cogan2024taboo,wilczewski2023language}. International students are constantly learning new skills to adapt to and operate effectively in unfamiliar environments~\cite{kim2015cross,lee2025into,ma2025exploring}.

The experiences of international students in the process of cross-cultural adaptation have attracted significant attention from various research communities such as Human-Computer Interaction (HCI), AI, and education~\cite{ma2025exploring, baines2022social,saha2024enhancing,arslan2025ai,serverovich2024role}. 
Many of these challenges are invisible, but persistent, which impede their ability to adequately learn about the host culture and find ways to adapt~\cite{sabie2022decade,bethel2020cross}
While universities offer support systems, such as international student services programs, student clubs or organizations, social events, advisors, friends, etc., these resources are often fragmented and may not be available 24/7 to address a student's immediate, day-to-day problems.

With the rise of conversational AI over the past few years, these tools have become nearly ubiquitous and are quickly becoming crucial tools in many domains, such as school~\cite{ammari2025students}, migration contexts~\cite{lee2025into}, mental health~\cite{lee2025artificial,pozzi2025keeping}, etc. 
Particularly, research has explored general student interactions with conversational AI, but much less is known about how \textit{international students} engage with these tools in depth~\cite{cena2026studying,ammari2025students}.
Although previous work has focused on understanding the intersection of AI and international students' experiences (e.g.,~\cite{ma2025exploring,wang2023exploring}), there remains a lack of user-centered research that provides in-depth insights into how AI fits into these students' unique life experiences. Research has yet to examine the role of AI not only through quantitative methods but also through qualitative approaches to uncover the nuances of incorporating AI into the international student ecosystem.
Thus, it is crucial to understand how AI can help mitigate the challenges international students face.

Through a mixed-method approach, we conducted a survey and interview studies to explore conversational AI usage by international students in the US, and their insights on the impact and potential of AI as a support tool tailored to their needs. First, we conducted a survey study completed by 60 international students in the US, addressing the following research questions: \textbf{(SRQ1)} For what types of challenges do international students use conversational AI tools during their adaptation process? \textbf{(SRQ2) }How do international students use conversational AI during their adaptation process? 
And \textbf{(SRQ3) }How do the students perceive the helpfulness of current conversational AI tools?

Our survey findings revealed that international students adopt distinct AI usage patterns for the different types of challenges they face: AI is preferred for functional, task-oriented support (i.e., short-term use) over emotional or longitudinal assistance (i.e., long-term use).
While students perceive AI as a powerful tool for navigating the immediate, practical challenges of living and studying in the U.S., they continue to prioritize human and experiential support for the deeper, more nuanced aspects of adaptation.

Guided by these findings, we conducted in-depth semi-structured interviews with 14 international students, examining the following research questions: 
\textbf{(IRQ1)} What draws international students to AI use or to avoid using AI?
\textbf{(IRQ2) }Why is AI used more for short-term issues than deeper, long-term challenges? And \textbf{(IRQ3) }What is required in an AI-powered support tool tailored to international students, especially one with more application for long-term support, if at all desirable?

Our interview findings reveal explanations for patterns observed in the survey. Conversational AI is used to handle the sudden increase in responsibilities as international students acclimate to a new country. 
However, students have major concerns about data privacy, especially when data can affect visa/immigration status, as well as the AI’s lack of social nuance. 
In addition to the concerns and limitations highlighted in our interviews, findings suggest that international students are interested in AI-powered support tools with more capabilities for long-term use, such as integration of AI-powered platforms in the current university support systems, as well as developing a tool that enables international students to find and sustain a sense of community.

The primary contribution of our paper is survey and interview findings on AI usage patterns by international students in the US across different challenge domains. These insights reveal how and why international students tend to use conversational AI to address their unique challenges. 
The second contribution of our work is a set of design ideas and recommendations by our participants for creating improved AI-powered support tailored to the unique needs of international students.


\section{Background and Related Work}

\subsection{International Students’ Cross-Cultural Adaptation}

Migration has expanded because of globalization, affecting hundreds of millions of people from diverse backgrounds, including international students~\cite{bethel2020cross}. 
Supporting migrants’ adaptation is crucial for both migrants and host communities. Cultural adaptation is a complex process where individuals aim to adapt to the host society’s cultural norms, interpersonal interactions, daily life logistics, and social practices while maintaining aspects of their own cultural identity~\cite{bethel2020cross, lee2025into,ma2025exploring}.

According to Kim's framework, Cross-Cultural Adaptation (CCA) is a dynamic process of ``stress-adaptation-growth,'' where the challenges of cultural shock act as a catalyst for personal transformation \cite{kim2015cross}. Kim’s framework argues that the experienced stress is primarily a prerequisite for adapting and growing in the new environment. 
This process forces individuals to constantly learn new skills to adapt and operate effectively in unfamiliar environments~\cite{kim2015cross,lee2025into}. Consequently, adaptation becomes an iterative process to resolve tensions between differing cultural perspectives and worldviews, and to manage the stresses associated with significant changes in daily life~\cite{bethel2020cross}.

Crossing cultures can be particularly stressful for international students, given the simultaneous need to adapt to multiple contexts; beyond crossing national and cultural boundaries, they may also experience transitions between educational systems grounded in different values and assumptions~\cite{bethel2020cross,ma2025exploring}. In general, over the past two decades, research has consistently shown that international students face four broad categories of challenges in their host countries: sociocultural, academic, psychological, and economic. In a 21-year trend analysis of 175 studies, Oduwaye et al.\cite{oduwaye2023trend} showed that across 2002-2022, there has been no real improvement in the types of challenges reported. This means that the same issues keep showing up, such as language, isolation, discrimination, academic stress, financial pressure, etc. Although challenges are similar globally, specifics vary by host country. 

For example, in a study of Asian international students in Scotland, Cogan et al.~\cite{cogan2024taboo} found that cross-cultural adaptation challenges can manifest as loneliness, communication barriers, and clashes over different cultural expectations of well-being in the host country~\cite{cogan2024taboo}. Furthermore, these adaptation challenges influence international students’ mental health and are intertwined with how cultural values shape not only students’ adaptation processes but also their ability to access support. Therefore, there is a need for appropriate services to ease adaptation stress for international students and promote healthier adaptation outcomes~\cite{cogan2024taboo}. 

To deal with this unique blend of challenges, international students turn to social and institutional support to cope. Social support includes friends and family from back home, friends and peers in the host country, and faculty advisors at the student's university, while institutional support includes international student services, mental health counseling, and other resources offered by the student's university. A 2020 survey of 256 international students,~\cite{Shu_Ahmed_Pickett_Ayman_McAbee_2020} found that perceived social and institutional support are strong predictors for successful cross-cultural adaptation, with social support from host-national friends (i.e., friends who are citizens of the host country) being the most helpful. In practice, many international students turn to family and friends from back home, which can have a potentially negative impact on adaptation, despite relieving the student's loneliness~\cite{Zheng_Ishii_2023}. Students also commonly seek community with other international students from the same country when possible to alleviate loneliness ~\cite{Cao_Zhu_Meng_2021}. Institutional support is more helpful for school-related cultural adaptation, such as gaining more fluency and confidence in English ~\cite{Shu_Ahmed_Pickett_Ayman_McAbee_2020}.

While bodies of work demonstrate the challenges international students face, and the support ecosystems they have access to, it remains unclear when students decide to use AI instead of, or in addition to, human support, and how they draw boundaries around appropriate AI use in certain domains.
Given the prevalence of AI tools, particularly conversational AI models such as ChatGPT, we aim to explore the use cases and impact of AI-as-a-service for international students’ needs.




\subsection{Perceptions Towards Conversational AI and Opportunities}

Previous HCI studies have advanced the understanding of public perceptions of conversational AI tools. For example, people’s perceptions and expectations of these non-human agents can be influenced by the design features of the conversational AI tools, the unique characteristics of individual users, and the contexts and scenarios in which conversational AI was deployed (e.g., ~\cite{fortunati2022people}).
There is evidence that users view conversational AIs, such as voice assistants, as tools rather than as human-like. 
An early study by Clark et al.~\cite{clark2019makes} in 2019 suggested a marked difference in the way participants discussed having conversations with agents compared to conversations with other people. In particular, conversations with agents were consistently described in functional terms, underscoring the utilitarian nature of these interactions rather than an emotional companion. Furthermore, the authors argue that we should view human-agent conversation as a new genre with its own rules, rather than simply as an imitation or substitute for human-human conversation.

However, it is important to consider that users’ perceptions can differ across user groups, disciplines, cultures, etc.~\cite{fortunati2022people, english2025rather,lee2025into}.
Through a comprehensive computational analysis of nearly one million social media posts related to conversational AI agents on Twitter and Weibo, Liu et al. found that people in the US and China differ in both the topics they discuss and the perceptions they hold toward conversational AIs~\cite{liu2024understanding}. Beyond cultural nuances in perception, other structural factors, such as political and economic contexts, also shape how AI tools are discussed and perceived~\cite{liu2024understanding}. Following differences in perceptions, AI adoption also differs; for example, Liu et al.’s analysis highlighted that Chinese users tended to approach AI tools in a more social and emotionally expressive way, whereas US users adopted a more task-focused approach in their interactions~\cite{liu2024understanding}.

\subsection{Conversational AI And Migrants}

Attitudes towards AI have been explored across different populations, including migrants living in a country. With the accelerated pace of globalization in modern society, migration has become a common phenomenon, driven by factors such as the pursuit of new experiences, higher education, career goals and requirements, global safety concerns, economic uncertainty, and family circumstances seeking better living conditions. 
Large Language Models (LLMs) have been actively explored to develop systems that reflect the diverse characteristics of marginalized user populations, facilitating effective communication between people and AI, and mitigating challenges
faced by migrants (e.g., ~\cite{lee2025into,tseng2023understanding}).

As Pozzi and De Proost~\cite{pozzi2025keeping} emphasize, when marginalized populations are excluded from design and evaluation processes, technologies may reproduce the very structural inequities they claim to alleviate. Such technologies may impose Western-centric solutions that fail to account for the unique bureaucratic and cultural barriers their users face~\cite{nourian2025cultural,nourian2025invest}. Consequently, without a user-centered research approach, AI-powered support risks further isolating international students by offering solutions that are functionally or culturally inappropriate.

The Human-Computer Interaction (HCI) community has contributed to understanding the experiences of marginalized populations, such as immigrants and refugees, and to designing technologies that help them overcome these challenges and adapt to their new communities (e.g.,~\cite{lee2025into,tseng2023understanding}).
Thus, when designing for immigrants, the specific situational contexts, communication practices, and habits of immigrants need to be considered (e.g., ~\cite{lee2025into,tseng2023understanding,bhandari2022multi}).

Recent work highlights a shift toward using LLMs as tools for identity negotiation and cultural adaptation.
For example, Lee et al. ~\cite{lee2025into} showed that migrants in Finland view conversational AI as a tool to support identity negotiation, cultural connection, and self-reflection during cultural adaptation.
Similarly, AI-mediated support can help empower migrants' autonomy by acting as a bridge between newcomer and host communities. For example, migrant workers have used AI to navigate the healthcare system, overcoming language and cultural barriers that frequently impede their access to and understanding of information in their host countries~\cite{tseng2023understanding}. 
Migrants often struggle to understand how social and logistical service settings work in the host country~\cite{lee2025into,Truong2024enhancing}. Moreover, their initial perceptions of host countries are usually shaped by media, research, or personal anticipation~\cite{lee2025into}. 

However, there are common concerns about AI-powered technology, including bias and stereotypical representations~\cite{lee2025into}, and whether AI models use trustworthy sources for their responses~\cite{Truong2024enhancing}. Beyond these ethical considerations, AI outputs should align with the mental models of its target users, a contextualization that is often absent in migration contexts. Research shows a fundamental divergence in how migrants and institutional advisors conceptualize the settlement process: migrants may often cluster questions and answers by everyday topics (e.g., banking, transport, healthcare) and use descriptive language to remember details, while advisors view the process sequentially (e.g., get residence permit, then bank account, next other services), focusing on steps and official requirements~\cite{Truong2024enhancing}. These different ``mental models'' may cause misunderstandings if they continue to persist in the AI technologies. 

\subsection{Conversational AI And International Students}
Conversational AI tools serve as learning tools for students, often assuming roles such as explainer, tutor, writing coach, job-search assistant, and emotional-support resource \cite{ammari2025students,zheng2025students,galdames2024impact}.  One common use case is information-seeking, retrieving factual information, clarifying concepts, and asking specific questions about academic, professional, and cultural topics \cite{ammari2025students}. Students also use AI for content generation, such as drafting essays, writing code, and writing resumes. 

Moreover, conversational AI is increasingly used for emotional support. AI-powered services can provide immediate, easily accessible mental health and counseling resources, as well as coping strategies, thereby contributing to emotional well-being and social interactions \cite{dhiman2025ai}. However, there remain ethical concerns about the lack of human empathy and potential for AI-generated responses to feel rigid or generic \cite{albikawi2025nursing}. 

International students often encounter significant difficulties adapting to new academic environments, managing cultural differences, and overcoming language barriers \cite{lund2025ai, dhiman2025ai, ma2025exploring, ittefaq2025factors}. In the US, many may come from non-English-speaking backgrounds and struggle with writing, grammar, and punctuation, and may receive limited informational support from peers and teachers \cite{ittefaq2025factors}. International students often struggle with writing, grammar, and punctuation in English \cite{ittefaq2025factors}. For instance, Chinese international students studying at American institutions have historically faced linguistic adjustment challenges, particularly with spoken English, largely due to educational practices that place greater emphasis on English grammar, reading, and writing in their home country \cite{zhang2018chinese}. This self-perceived linguistic inadequacy can hinder class participation and academic success, leading to challenges with psychological well-being and transnational adjustment \cite{zhang2018chinese}. These initial obstacles, coupled with cultural adaptation difficulties and logistical issues like finding accommodation, contribute to stress, depression, and loneliness \cite{saha2024enhancing, ma2025exploring}.


Moving to a foreign educational environment introduces unique challenges, including linguistic barriers, cultural adaptation challenges, psychological stress, and unfamiliar academic expectations \cite{ittefaq2025factors, saha2024enhancing, arslan2025ai}. Conversational AI technologies are increasingly leveraged to address these systemic issues, offering personalized, real-time support that traditional institutional mechanisms often fail to provide consistently \cite{arslan2025ai, saha2024enhancing}.

AI applications that promote cross-cultural communication and understanding help create a comprehensive and supportive environment for international students \cite{dhiman2025ai}. These tools provide support previously limited by language challenges, offering individualized learning, real-time writing assistance, and rapid access to knowledge \cite{ittefaq2025factors}. Postgraduate international students, for example, have used AI tools like ChatGPT to simplify complex academic jargon or translate written paragraphs between languages (e.g., French to English, English to Chinese) \cite{arslan2025ai}.


\section{Online Survey Method}
We ran an IRB-approved online survey with international students enrolled in US institutions to explore their use and adoption of conversational AI tools during their adaptation to life in the US Our survey questions were framed using Kim’s stress–adaptation–growth model~\cite{kim2015cross}.

\subsection{Survey Materials and Procedure}
We distributed our survey to current international students aged 18 or older at US institutions.
To avoid complications in the data, we recruited only first-time international students who had not studied in another country before the United States. Our justification was to maintain a homogeneous baseline of participants and focus specifically on the experiences of those moving directly from their home culture to the US. This exclusion criterion ensures that our data on domestic challenges and AI usage patterns were not impacted by coping mechanisms developed in other international contexts.
Our survey was advertised on multiple platforms, including LinkedIn, Twitter, and Reddit, as well as through our professional network.
In addition to social media pipelines, we identified relevant groups and mailing lists of institutions in the US and reached out to the primary contact person (e.g., department points of contact or instructors) to share our study with their students. 

With 30 questions across four parts, our survey was designed to capture international students’ demographic information, stress and challenges, and experiences with conversational AI tools during their cultural adaptation process.

\textbf{Part 1: Demographic and Background Information:} This survey section collected participants’ general background information, including age, gender, country or region of origin, academic level (undergraduate, master’s, or doctoral), program of study, length of stay in the United States, etc. Participants were also asked about their prior experience with conversational AI tools, including their digital skills and the duration of their use, to provide context for interpreting their responses in later sections.

\textbf{Part 2: Stress and Adaptation Challenges: }
This survey section focuses on the first phase of Kim's framework, collecting data on the challenges that international students frequently encounter. 
We included two questions in this part to capture the challenges participants have faced in adapting to life in the US and the kinds of support they seek. The first question was an open-ended item asking students about the biggest challenges or sources of stress they faced while adjusting to life in the US. We also provided examples to clarify the question (e.g., language, making friends, communicating with professors, managing stress, understanding their studies, etc.). 
Next, we included a multiple-choice question asking students what sources of support they prefer when facing challenges.
The goal of this section was to contextualize the challenges and adaptation experiences to prepare participants for the next parts of the survey.


\textbf{Part 3: Use of AI Tools for Adaptation and the Current Patterns: }
Next, our survey shifted focus to the adaptation phase of Kim's framework to examine students' actions toward cultural adaptation in our study. We wanted to know how students use AI in this process.
First, participants reported on their engagement with conversational AI tools (e.g., ChatGPT, Gemini) and the challenges they addressed using them. Questions included the types of conversational AI tools used and their purposes (e.g., socio-cultural challenges, academic challenges), frequency of use, and preferred prompting language (e.g., English or native language). 
We included multiple options for the challenges AI is used to address. The choices were derived from prior work that identified the most prominent stressors international students encounter~\cite{oduwaye2023trend}.
Follow-up questions were split into four matrices based on students' choices. For example, if a student chose an option related to academic challenges, a 5-point Likert-scale matrix was displayed to elicit feedback on using AI for academic challenges.
At the end of Part 3, participants were asked to answer three 5-point Likert-scale questions to provide general feedback on the perceived AI's usefulness in addressing their challenges, their adaptation process, and their growth and improvement. In addition, some open-ended items invited participants to elaborate on how AI tools have helped or failed to support their needs.

\textbf{Part 4: Optional Reflection: }
The final section included optional open-ended questions that invited participants to share any additional thoughts, personal experiences, or reflections on the use of conversational AI tools in their adaptation journey. 

\subsection{Survey Participants}
Our survey received 194 entries; after removing duplicates, invalid, and incomplete entries, we ended up with 60 responses in total. Invalid entries were the ones that did not meet our criteria (e.g., first-time international students residing in the US). 
We also considered responses incomplete if the respondent failed to answer at least 65\% of the questions. These entries were excluded because they did not provide full responses to the critical survey sections that were necessary for our primary analysis, despite missing demographic, open-ended, or optional reflection questions.

Out of 60 responses, 56 reported using conversational AI to support their cultural adaptation process in the United States.  The participants included 33 women and 27 men, with an average age of 26. The majority were graduate students: 34 PhD students, 16 Master's students, and nine undergraduate students. The majority of participants rated themselves as having good or excellent skills in both digital literacy and English fluency. Forty-six participants reported using English exclusively when interacting with conversational AI, whereas only two used their native language. The most commonly used conversational AI was ChatGPT, used by all participants who had used AI; the second-most common was Gemini, with 32 participants reporting they had used it.


\subsection{Survey Analysis}


We first read and reviewed all responses to ensure data quality and establish a holistic understanding of the participants' feedback. This process of familiarization helped contextualize the ratings and qualitative components. In addition, all participants were assigned anonymized IDs, such as S1, S2, etc.
Our analysis focused on answering three research questions: \textbf{(SRQ1)} For what types of challenges do international students use conversational AI during their adaptation process? \textbf{(SRQ2)} How do international students use conversational AI during their adaptation
process? and \textbf{(SRQ3)} How do international students perceive the helpfulness of current conversational AI tools? 

We tested the following hypothesis based on prior research \cite{gajos2022people, karny2024learning}. \textit{Students likely use AI for their adaptation challenges and turn to it when they face an immediate, task-oriented problem (such as an assignment, email, or language query), rather than when they face a chronic, in-depth challenge, such as loneliness.}

\subsubsection{Descriptive Statistics, and Statistical Analysis}

Our team started the quantitative analysis with descriptive analysis to summarize the dataset. 
Means, medians, percentages, and standard deviations were calculated for 5-point Likert scale questions. The original scale was Strongly Agree - Strongly Disagree, which was converted to a 1-5 scale for calculations. This was done across all Likert-scale questions in the survey.

Next, we used correlation techniques to assess the significance and direction of linear relationships among variables in our survey. We used Spearman's Rank Correlation Coefficient, Mann-Whitney U test, and Wilcoxon signed-rank Test.
For example, we generated correlation matrices to investigate relationships between variables. This nonparametric approach was selected because the Likert-scale data are ordinal.
Based on the correlation results, a set of nonparametric tests was conducted to assess the statistical significance of key findings from the descriptive analysis.







\subsubsection{Thematic Analysis}
For the open-ended questions, thematic coding was performed. Our analysis focused on qualitative interpretation of open-ended responses to contextualize our findings from the quantitative data. 

One researcher reviewed the raw responses from open-ended questions to immerse themselves in the data. The researcher generated initial codes to capture specific participant sentiments regarding AI usage and adaptation challenges.
The initial codes were then collated into categories to identify international students' challenges, practical needs, and patterns of AI use.
We continually checked emerging themes against the quantitative results to ensure that the qualitative interpretations accurately captured the survey findings.

\section{Survey Findings}


Our results outline the relationship between the challenges faced by international students and their subsequent usage of conversational AI tools.  AI adoption is concentrated in high-stakes, task-oriented domains such as academic writing and logistics, despite students reporting a wide range of socio-cultural and psychological challenges. Although students still choose human-centered help for long-term emotional and cultural adaptation, they prioritize AI for its on-demand nature. 
Please see Appendix~\ref{apendix} for more detailed tables on survey data.

\subsection{Cultural Adaptation challenges international students face, and the role of AI in resolving them}
Overall, survey responses indicate that students experience multiple challenges, and some domains are more common than others.
To learn further, we asked students to respond to an open-ended question. 56  out of 60 respondents answered this question: \textit{As an international student, what are the biggest challenges or sources of stress for you while adjusting to life in the U.S.? (For example: language, making friends, communicating with professors, managing stress, or understanding your studies.)}

Socio-cultural, language, and communication-related challenges emerged as the most frequently reported, with 75\% of students reporting difficulties in these areas. Academic challenges were the second most commonly reported, with 43\% of students noting issues with coursework, academic expectations, or studying in a new educational system. Challenges related to managing physical or mental health, stress, or homesickness (marked as psychological challenges) were also present for 39\% of the students. In contrast, logistical challenges were mentioned less often than other domains: 29\% of students reported difficulties with tasks such as navigating legal or administrative processes (e.g., filing taxes). In general, these findings suggest that logistical issues are less frequent; socio-cultural and language-related challenges are the most common sources of stress for international students in our sample.

\begin{figure}
    \centering
    \includegraphics[width=1\linewidth]{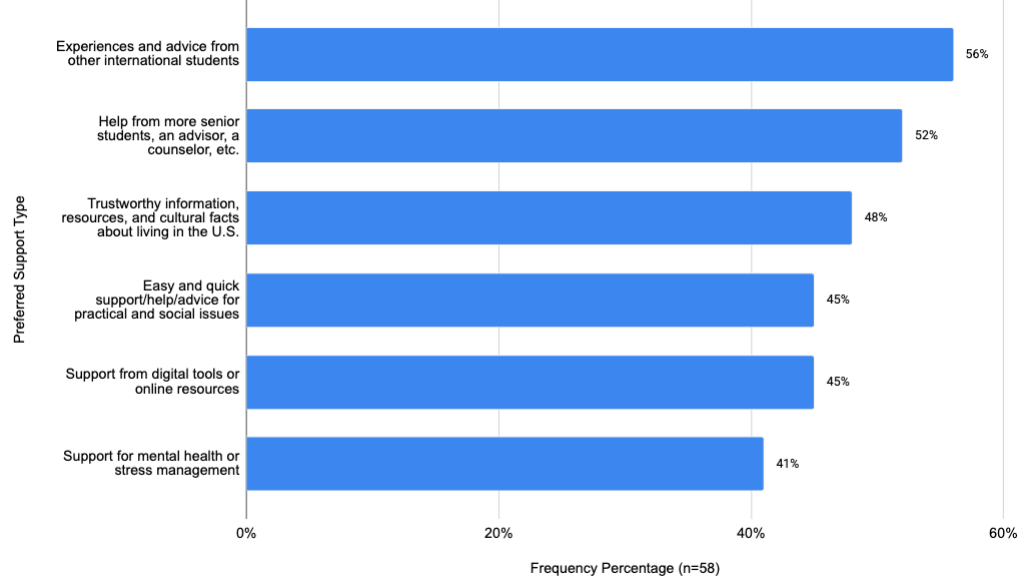}
    \caption{Support types that international students prefer}\label{support-type}
        \Description{}
\end{figure}

Following up on the students' challenges, the survey also asked about the types of support they prefer for those challenges. Findings indicate that the most preferred support is human-based and experiential (e.g., advice from others and help from seniors/advisors). Support from digital tools or online resources was selected as the preferred option for stressors that require quick solutions (Figure ~\ref{support-type}).

\subsection{Patterns of AI usage for navigating adaptation challenges}

In this stage of analysis, we tested our hypothesis: Students likely use AI for their adaptation challenges and turn to it when they face an immediate, task-oriented problem (such as an assignment, email, or language query), rather than when they face a chronic, in-depth challenge, such as loneliness. To test this hypothesis, our survey combines Challenge Prevalence questions with AI Usage questions.
Out of 60 responses, four students indicated that they have never used conversational AI as a support tool for their adaptation challenges stating, they ``never thought to'' (S23), or ``felt it was wrong and ultimately wouldn't benefit from the use'' (S20).

\begin{figure}
    \centering
    \includegraphics[width=0.6\linewidth]{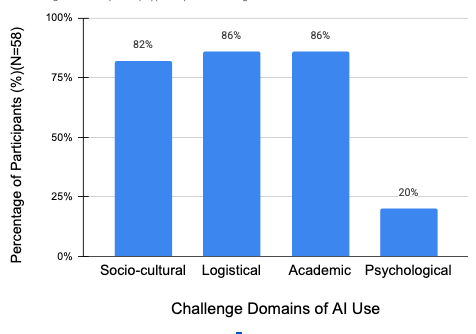}
    \caption{AI usage distribution among the four challenge domains}
        \Description{}
        \label{AI-use}
\end{figure}

\begin{figure}
    \centering
    \includegraphics[width=1\linewidth]{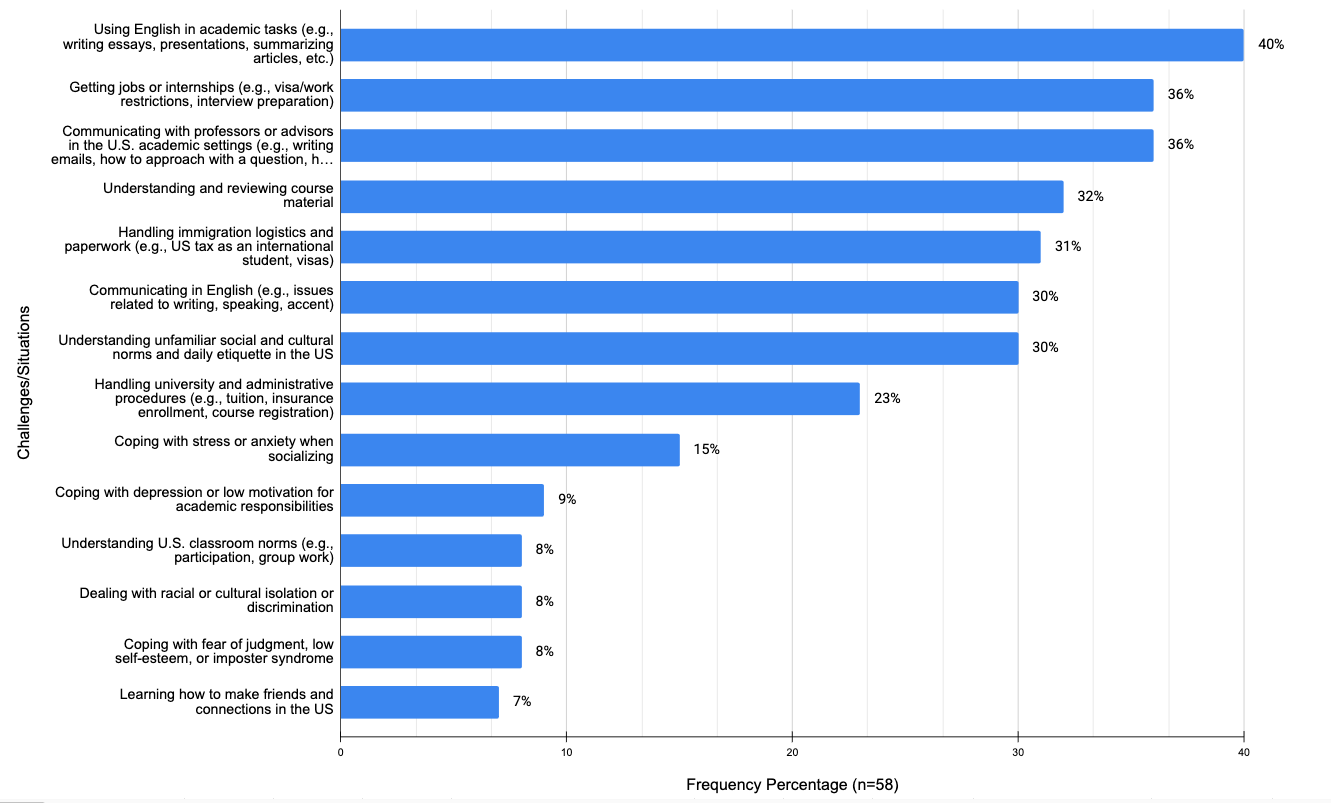}
    \caption{AI usage distribution for detailed tasks among the four challenge domains} \label{domain-tasks}
        \Description{}
\end{figure}

Findings show that AI was used most frequently for logistical and academic challenges (See Figure \ref{AI-use}). Out of 56 AI users (from the previous question),  48 (86\%) of students reported using AI to help with tasks such as using English in academic settings (e.g., writing essays or communicating with professors), handling visa restrictions, immigration logistics, or paperwork. (See Figure \ref{domain-tasks}). Socio-cultural challenges were the next most common domain, with 46  (82\%) of students indicating that they use AI to address needs such as understanding unfamiliar social and cultural etiquette or issues related to writing, speaking, or accent. In contrast, only 11 (20\%) of students reported using AI to address psychological or mental health-related challenges, such as stress, loneliness, or homesickness. The lack of use for psychological challenges shows a sharp drop-off when the challenge moves from a task (essay) to a chronic state (loneliness).

\subsubsection{AI as a tool for quick, short-term, and task-oriented support}

\begin{figure}
    \centering
    \includegraphics[width=1\linewidth]{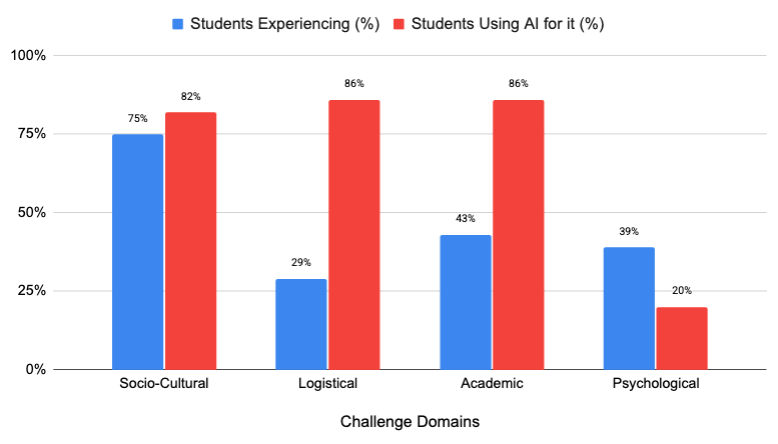}
    \caption{Comparative analysis of international students' challenges and usage of conversational AI for each challenge domains}
        \Description{}
        \label{challenge-use-side-by-side}
\end{figure}

Comparing the challenges students indicated and the use of AI, a notable mismatch appears in the domain of logistical challenges; although only 29\% of students reported experiencing logistical difficulties, 86\% reported using AI for logistical tasks, making it the most common domain of AI support. See Figure \ref{challenge-use-side-by-side}.
The examples we collected shows that students rely on AI to interpret complex rules: ``\textit{It mostly helped with practical stuff like opening a bank account, explaining what certain forms are for [...] and so}.'' (S12) or translate bureaucratic language: ``\textit{I used AI to know more about bureaucratic stuff like visa forms.}'' (S30)

At the same time, the findings show a strong alignment between some of the most common challenges and students’ AI use: Socio-cultural challenges were the most frequently reported (75\%), and AI is also widely used to address them (82\%). Similarly, academic challenges were the second most common challenge (43\%), and AI use for academic support was comparably high (86\%).
Students who reported socio-cultural and academic challenges often described AI as a language assistant or editor, helping them prepare emails, clarify social norms, and understand study material.

Regarding the use of AI for psychological challenges, despite 39\% of participants reporting them, only 20\% use AI as a support tool, the lowest usage rate across all domains. The open-ended responses suggest that when students do involve AI in this domain, they typically frame it in task-oriented, advice-seeking, or self-management formats. For example, one student highlighted: ``\textit{I have asked AI several times if I am doing enough to pursue the future career I am aiming for.}'' (S1), and another turned to AI to cope with stress and low motivation by organizing their workload: ``\textit{During stressful weeks, I’ve used AI to get tips on managing study pressure or organizing my schedule better. It helped me feel more in control and focused.}'' (S35). However, several participants explicitly said that they do not use AI for psychological challenges, even when they reported struggling with issues such as stress or anxiety: ``\textit{I never used it (AI) for this (psychological challenges).}'' (S25). These patterns help explain that despite the existence of stress and homesickness in students’ narratives, AI is not considered a frequent resource for managing psychological and chronic emotional well-being. In general, these findings suggest that students primarily engage with AI as a tool for information-seeking and task-oriented support (logistical, academic, and socio-cultural), rather than as a resource for managing psychological, emotional, or chronic matters.

One of the most prominent reasons international students turn to AI for cultural adaptation challenges is its on-demand, quick nature. Students wrote that AI has become their first option due to the urgency of their tasks. For example, S14 described AI like a first-aid: ``\textit{The go-to place to get quick, trustworthy, and to the point information or solution to a problem. Its like a first-aid}'' (S14)

Furthermore, the on-demand, easy access to AI has been a factor, especially when human support is not available at all times. S29 emphasized AI helping with filling the gap of human capacity to take on other people's venting: 
\begin{quote}
    ``\textit{For other problems, if I were to go to a friend, they need to be available and have the emotional capacity to share my baggage as well - which you can't expect everyone to do whenever you run into issues}'' (S29)
\end{quote}

\subsection{AI's helpfulness for dealing with cultural adaptation stressors, the adaptation process, and growth}

Looking at the data on three variables of ``perceived usefulness of AI responses'', ``students' satisfaction'', and ``their intent to continue using AI'' for each challenge domain (i.e., matrices), we attempted to better understand the uncovered mismatch and alignments between experienced challenges and AI usage. These were asked through 3 Likert scale questions for each challenge domain, where students rated their experiences from 1 (Strongly Disagree) to 5 (Strongly Agree). For example, 
\begin{enumerate}
    \item  Usefulness: AI provided useful solutions to address my challenges
    \item  Satisfaction: I am satisfied with the help I received from AI for my challenges
    \item  Intent to Continue: I would like to continue using AI for challenges
\end{enumerate}

For all three variables, the median is 4, indicating that the majority of students agree that using AI is satisfactory and useful and are willing to continue using it. At least 50\% of our respondents chose Agree or Strongly Agree for all metrics, showing a high baseline of AI adoption. There is a hierarchy of how international students perceive AI's value. Academic challenges had the highest scores across all three metrics (e.g., Usefulness: Mean=4.27, SD=0.49, Satisfaction: Mean=3.88, SD=0.67, and Intent to Continue: Mean=4.38, SD=0.49). These findings suggest that AI has become an integrated, non-negotiable part of the international student academic experience.  For academic challenges, AI is seen as highly reliable for hard skills (grammar, structure, and summarizing).

Logistical was the second-highest rated domain (Usefulness: Mean=4.04, SD=0.56, Satisfaction: Mean=3.78, SD=0.76, Intent to Continue: Mean=4.22, SD=0.64). Students find AI highly useful for navigating self-reliance. It acts as a 24/7 consultant for complex systems like US immigration and taxes. 

Socio-cultural shows a notable drop in satisfaction, with the lowest satisfaction mean across the entire table (Usefulness: Mean=3.8, SD=0.72, Satisfaction: Mean=3.49, SD=0.66, Intent to Continue: Mean=3.8, SD=0.92). There is not much satisfaction, but higher usefulness. However, students still want to continue using AI anyway. Given the massive need in this area (75\%), it seems students have to use AI even if they are not highly satisfied. AI's help is ``just okay''. It seems like AI is still liked, but not as good as other logistical and academic support. 

Psychological challenges present the most significant mismatch in the data. It has the lowest (Usefulness: Mean=3.63, SD=0.8) but a high (Intent to Continue: Mean=4, SD=1). It suggests that even though students know AI is not great at ``emotional'' support, they still plan to use it. Students choose not to use AI as their go-to support for psycho (20\% use it). The Satisfaction (Mean=3.55, SD=0.93) and especially the usefulness mean rates show that students do not see AI as the best support for this job. Since the preferred support type is human-based (peers, counselor, etc.), and also, psycho needs are chronic, students don’t turn to it. Whenever they did, they were not highly satisfied with its help. Students explicitly rate AI as less useful here than in any other category. See table ~\ref{intent-use-satis} in Appendix for a full breakdown. This is where the Kruskal-Wallis test ($p=0.002$) in the following section highlights the most significant difference.


\subsubsection{AI as a practical tool for short-term support}
The results from the Kruskal-Wallis Test on the three variables (perceived usefulness, satisfaction, and intent to continue using AI), indicated that there is a significant difference between usefulness ($\chi^2$ = 15.14, p-value=0.002) and intent of continuation ($\chi^2$ = 11.27, p-value=0.010), but not for the satisfaction variable ($\chi^2$ = 6.74, p-value=0.08). This shows that AI's usefulness and the intent to continue using it (a student's future actions) differ significantly across challenge domains. Such significance demonstrates that students indeed treat and perceive AI differently across challenge categories.

A follow-up post hoc pairwise test was then conducted for usefulness and Intent of Continuation, revealing where the differences lie across domains. Findings from the post hoc tests indicate that Academic category is the clear "winner" in terms of practical usefulness. Usefulness was significantly higher in Academic vs. Socio-cultural: this means that students find AI more useful for Academic tasks than for Socio-cultural tasks. Students are significantly more likely to continue using AI for Academic tasks (p=0.0054) because it is significantly more useful for those tasks (p=0.0048). Students find AI significantly more useful for academic tasks than for navigating social or cultural norms.

Logistical vs. Academic was not significant ($p = 0.28$). Students view AI as being similarly high in usefulness for both school and administrative survival (visas, taxes). Sociocultural vs. Psychological yielded a ($p = 1$). Students view AI’s usefulness for social norms and mental health as virtually identical, and both are equally lower than the academic/logistical scores. Academic vs. Psychological is not significant here ($p = 1$), even though it was significant for usefulness. Even though students find AI less useful for psychological needs, their intent to keep using it is high enough that it doesn't statistically differ from their intent to use it for school. This points to a lack of other options—students will keep using AI for stress management simply because it is accessible. The difference in student behavior (continuation) is a direct response to the significant difference in the AI's usefulness within the 4 challenge categories.

Following up with a Wilcoxon Signed-Rank Test, we compared students’ ratings of Satisfaction and Usefulness with their intention to continue using AI within each challenge domain. We sought to determine whether there is a significant gap between the scores, i.e., whether Intent to Continue scores consistently exceed Satisfaction and Usefulness across domains. A primary finding across all four domains is the statistically significant difference between current Satisfaction and the Intent to Continue using AI. Socio-cultural ($p=0.0106$), Logistical ($p=0.0001$), Academic ($p<0.001$), and Psychological ($p=0.0253$) (See table \ref{wilcoxin-signed-rank}. While medians for both variables often sit at 4.0 (Agree), the distributions differ significantly. This suggests that a student's immediate positive feeling (satisfaction) is not a perfect predictor of future behavior. Factors driving transactional satisfaction (e.g., a quick answer for a paper) are distinct from those driving sustained integration. In all domains, Continuation scores were significantly higher than Satisfaction scores, indicating that students plan to keep using AI because they feel they need to, even if they aren't fully satisfied with the experience. Moreover, we find no statistical difference between students' ratings of the usefulness of AI responses and their intent to continue using AI across the three challenge domains of socio-cultural, logistical, and academic. Since the two variables are closely linked, it is most likely that students will continue using AI for those three challenges if they find it useful.

\begin{table}[H]
\caption{Wilcoxon Signed-Rank Test for all four domains for Satisfaction/Intent to Continue AND Usefulness/Intent to Continue}\label{wilcoxin-signed-rank}
\begin{tabular}{|l|l|r|c|}
\hline
\textbf{Domain}         & \textbf{Comparison}                    & \multicolumn{1}{l|}{\textbf{p-value}} & \multicolumn{1}{l|}{\textbf{Finding}} \\ \hline
\textbf{Socio-cultural} & \textbf{Satisfaction vs. Intent to Continue} & \textbf{0.0106}                       & \textbf{Significant}                  \\ \hline
Socio-cultural          & Usefulness vs. Intent to Continue            & 0.9827                                & Not Significant                       \\ \hline
\textbf{Logistical}     & \textbf{Satisfaction vs. Intent to Continue} & \textbf{0.0001}                       & \textbf{Significant}                  \\ \hline
Logistical              & Usefulness vs. Intent to Continue            & 0.0896                                & Not Significant                       \\ \hline
\textbf{Academic}       & \textbf{Satisfaction vs. Intent to Continue} & \textbf{0}                            & \textbf{Significant}                  \\ \hline
Academic                & Usefulness vs. Intent to Continue            & 0.1317                                & Not Significant                       \\ \hline
\textbf{Psychological}  & \textbf{Satisfaction vs. Intent to Continue} & \textbf{0.0253}                       & \textbf{Significant}                  \\ \hline
\textbf{Psychological}  & \textbf{Usefulness vs. Intent to Continue}   & \textbf{0.0455}                       & \textbf{Significant}                  \\ \hline
\end{tabular}
\end{table}

 Overall, international students had a positive view of conversational AI's helpfulness, as shown in Table \ref{global-q}, presented by the high percentage of positive responses (Agree/Strongly Agree) and a consistent median score of 4 (Agree) for all three of the global questions about helpfulness.

\begin{table}[]
\caption{International students' responses on AI's helpfulness for dealing with challenges, adaptation, and growth}\label{global-q}
\begin{tabular}{|l|c|c|c|c|}
\hline
\textbf{\begin{tabular}[c]{@{}l@{}}Perceived AI's Helpfulness for:\end{tabular}}   & \multicolumn{1}{l|}{\textbf{Mean}} & \multicolumn{1}{l|}{\textbf{Median}} & \multicolumn{1}{l|}{\textbf{\begin{tabular}[c]{@{}l@{}}Positive Counts \end{tabular}}} & \multicolumn{1}{l|}{\textbf{\%}}                                                                                                                          \\ \hline
\textbf{\begin{tabular}[c]{@{}l@{}}Dealing with challenges and stressful situations\end{tabular}} & 4.00                               & 4                                    & 45 out of 53                                                                                                   & 84.90\%                                                 \\ \hline
\textbf{\begin{tabular}[c]{@{}l@{}}Growing and Improving\end{tabular}}             & 3.79                               & 4                                    & 33 out of 52                                                                                                   & 63.46\%                           \\ \hline
\textbf{Adapting}                                                                     & 3.67                               & 4                                    & 41 out of 53                                                                                                   & 77.40\%                          \\ \hline
\end{tabular}
\end{table}


\subsection*{Summary of Findings}

In summary, our findings show that AI adoption among international students, is centered around functional, task-oriented support over emotional or long-term support. While socio-cultural and language barriers remain the most prevalent challenges, AI is mostly used for academic and logistical tasks, with usage rates (86\%) far exceeding the reported frequency of logistical difficulties (29\%). This logistical mismatch suggests that while students face fewer logistical difficulties, the high-stakes nature of these tasks makes AI's on-demand, first-aid capability invaluable. Statistical analysis confirms that students’ intent to continue using AI is driven primarily by its practical utility in academic and logistical domains rather than general feeling of satisfaction. However, despite the prevalence of psychological challenges like homesickness and loneliness, international students tend to set more strict boundaries in this domain, mostly using AI for time management rather than emotional support. Ultimately, while students perceive AI as a powerful tool for navigating the immediate, practical challenges of living and studying in the U.S., they continue to prioritize human and experiential support for the deeper, more nuanced aspects of adaptation.



\section{Interview Method}\label{interview-method}
We conducted follow-up interviews with international students, to explore the AI adoption patterns, students motivations, and boundaries in more depth. Our interview study procedure was approved by our Institutional Review Board (IRB).
\subsection{Participants}
\subsubsection{Recruitment Process}

At the end of the survey, participants could opt in to be contacted for future interview studies. Participants who opted in are the first group the team contacted for recruitment. Twenty-nine survey participants initially volunteered for an interview study. The team sent an invite to all volunteer participants, and ended up conducting interviews with 12 participants from the survey and two with external sources. One non-survey participant was referred to an interview, and another was recruited through the team's network. If students had previously completed the survey, they were asked to schedule an interview time directly. If they were external participants, they were provided with a screener to determine whether they qualified for the interview. 

\subsubsection{Demographics}

Participants were from diverse backgrounds and nationalities. Of the 14 participants, the majority reported being from India (n=6, 42.8\%). Other participants represented Nigeria, Colombia, Taiwan, Canada, Iran, Bangladesh, Myanmar, and China (1 each, respectively). For the gender distribution, seven participants identified as men (50\%), and seven identified as women (50\%). The average age of the participants was 25.33 years (SD=3.70), with a range from 19 to 32 years. 

Participants were both undergraduate and graduate students. Among all participants, six (42.8\%) were PhD students, six (42.8\%) were Master's students, and two participants (14.3\%) were undergraduate students. Their majors were Computer Science, Human-Computer Interaction (HCI), Computational Science, Information Science, Industrial Engineering, and Economics. 33.3\% were in their first year, 33.3\% in their second year, and the remaining 33.4\% were in their third year or beyond.

13 (92.8\%) participants reported being fluent/advanced in English, with one participant reporting intermediate proficiency. 
On average, participants had lived in the United States for two years, ranging from recent arrivals (two months) to longer residents (six years). 

We asked our participants to self-report their digital skills with AI defined as (``the ability to use conversational AI technologies and their information effectively''). Four (28.57\%) participants reported ``Excellent'', eight (57.14\%) reported ``Good'', and two (14.2\%) reported ``Average''. 
See Table \ref{demo-table} for detailed demographic information.

\begin{table}[h] 
\centering
\caption{Summary of participant demographics of the interview study($N=14$)}
\label{demo-table}
\begin{tabular}{@{}l|l|l@{}}
\toprule
\textbf{Category} & \textbf{Returning Participants ($n=12$)} & \textbf{New Participants ($n=2$)} \\ \midrule
\textbf{Gender} & 7 Man, 5 Woman & 2 Woman \\
\textbf{Age} & Mean: 25.3 ($SD=3.7$) & Mean: 25.5 ($SD=2.1$) \\
\textbf{Country of Origin} & India (4), Others\textsuperscript{a} (8) & India (2) \\
\textbf{Education Level} & & \\
\quad PhD & 6 (50.0\%) & 0 (0\%) \\
\quad Master's Degree & 4 (33.3\%) & 2 (100\%) \\
\quad Undergraduate & 2 (16.7\%) & 0 (0\%) \\
\textbf{English Proficiency} & & \\
\quad Fluent / Advanced & 11 (91.7\%) & 2 (100\%) \\
\quad Intermediate & 1 (8.3\%) & 0 (0\%) \\
\textbf{Avg. U.S. Residency} & 1.69 Years & 2.5 Years \\
\textbf{AI Digital Skills} & & \\
\quad Excellent & 2 (16.7\%) & 2 (100\%) \\
\quad Good & 8 (66.7\%) & 0 (0\%) \\
\quad Average & 2 (16.7\%) & 0 (0\%) \\ \bottomrule
\end{tabular}
\begin{flushleft}
\small \textsuperscript{a} Nigeria, Colombia, Taiwan, Canada, Iran, Bangladesh, Myanmar, and China ($n=1$ each).
\end{flushleft}
\end{table}

\subsection{Interview Materials and Procedure}
We conducted semi-structured interviews as a follow-up to the survey to better understand how and why international students use conversational AI during their adaptation process. The interview design followed an explanatory mixed-methods approach, where questions were directly informed by patterns and gaps identified in the survey results.

We asked participants to reflect on their recent or previous use of conversational AI. They were encouraged to keep track of situations in which they use AI before the session to ground their responses in concrete examples and experiences. 

All interviews were conducted on Zoom and lasted approximately 45–60 minutes. To begin the session, we explained the study's purpose to our participants and clarified that the interview was a follow-up to the survey. They were informed that the session would be recorded for transcription purposes, and their responses would remain anonymous. We collected participants' written consent before each interview.

The interviews followed a semi-structured guide that covered key questions while leaving room for relevant, unscripted follow-ups and for participants to elaborate on their experiences. The interviews were structured in three parts:

\textbf{Part 1:} First, participants were asked to describe their experiences adapting to life in the United States, focusing on specific challenges they encountered and how they responded to them. These questions aimed to establish context and highlight concrete examples of adaptation-related difficulties, such as navigating unfamiliar academic systems, social environments, or daily logistics. Follow-up questions were used when needed to encourage participants to reflect on specific moments or situations rather than general impressions.

\textbf{Part 2: }Next, the interview focused on participants’ use of conversational AI. We asked whether they had used AI tools to address their challenges and, if so, how and why they chose to use them. 
For participants who reported using AI, we asked about the types of tasks they used AI for, what made AI a suitable option compared to other support resources, and in which situations they chose not to use it. For participants who did not use AI, we explored their reasons for avoiding it and what alternatives they relied on instead.
To support this discussion, we provided participants with a brief reference summarizing four domains of challenges identified in the survey: academic, socio-cultural, logistical, and psychological. This was presented using simple slides during the interview to ensure a shared understanding of the domains without constraining participants’ responses. Participants were encouraged to relate their experiences to these categories but were not restricted to them.

\textbf{Part 3: }Finally, we focused on participants' perceptions of AI as a short-term or long-term support tool. We introduced this distinction explicitly through the slides and asked participants to reflect on whether they view AI as a tool for immediate, task-based needs or as a resource that could support ongoing challenges over time (i.e., long-term use). Follow-up questions explored why AI is primarily used the way it is, whether participants would want long-term support from AI, and what changes, considerations, requirements, etc. would be needed to make such use appropriate or desirable.

Throughout the interview, participants were encouraged to draw on their own experiences and, when helpful, refer to past interactions with AI tools (e.g., recalling prior prompts or use cases). However, they were not required to share any personal chat histories or their screen. 


\subsection{Interview Analysis Process}
The researchers led the data analysis using Braun and Clarke's reflexive thematic analysis approach~\cite{braun2021can} 
We wrote accurate transcripts, using the Zoom recordings, which also helped each researcher become familiar with the data. Next, we used Miro (\url{https://miro.com}) to support a remote, collaborative, and iterative process for discussing and analyzing interview data. Moreover, we assigned IDs to our interview participants using letter P (e.g., P1, P2) for anonymity.

Data analysis was conducted through an inductive approach. To enable an effective analysis, the team members split the interview data (transcripts) among themselves. 
Analysis began with all researchers identifying initial codes in their assigned transcripts and reviewing others' codes. Through iterative rounds of discussions, the team discussed the analysis procedure and identified patterns from the initial codes on Miro. Next, we collated the patterns and generated themes that addressed each of our research questions.

We do not report inter-rater reliability based on the guidelines for Reflexive Thematic Analysis (RTA) established by Braun and Clarke~\cite{braun2019reflecting, braun2021can}. In this method, rigor is established through transparency and reflexive engagement, rejecting coder agreement. Thus, we emphasize researchers' reflexivity, where personal or disciplinary backgrounds are acknowledged as shaping the interpretation of data. Given the qualitative nature of the second study, our goal was not to use interview data to generalize findings but to explore and describe human experiences in-depth~\cite{mcdonald2019reliability}.
%

\subsubsection{Positionality}\label{positionality}
Our research team consists of a diverse, multidisciplinary group of researchers with expertise in AI, Human-Computer Interaction (HCI), psychology, design, and accessibility. The lead researcher of this study is from the Global South and came to the US as an international student. She is experienced in conducting research with underrepresented communities from different societies in the AI and HCI research domain. Moreover, the broader team includes researchers from both the US and the Global North, one of whom moved to the US. These combinations have allowed for some varied perspectives on domestic institutional structures and international transitions.

\section{Interview Findings}

We identified three main themes that encompass in-depth insights on AI use cases by international students, external factors in participants' lives that encourage AI usage, and AI's limitations that draw students to use AI for more short-term, task-oriented resources rather than a long-term, continuous support over a period of time (IRQ1 and IRQ2). Section~\ref{design-ideas} continues reporting on the findings but focuses on the design solutions offered by our participants to address IRQ3.


\subsection{Theme 1: Cross-Domain Usage of AI Through The Whole Timeline of Migration}

Consistent with survey findings, interview data confirmed a pattern of usage across the four main challenge domains, particularly for situations where international students struggle to understand and solve unfamiliar circumstances ``\textit{that's unique to them}'' (P2). However, using AI also begins before physical arrival in the US, the host country. 

\subsubsection{Post-arrival use cases of AI}

Our participants raised socio-cultural challenges they face and emphasized the importance of proper communication with people in the US, whether with their peers, professors, or people outside the university. In some cases, participants use AI to improve their language skills and learn how to mitigate awkward moments when they experience cultural shock. P5 referred to the phrase ``\textit{in Rome, be a Roman}'' to emphasize the importance of cultural adaptation in the host country.

Beyond navigating social and cultural norms, participants extended their AI use into their academic lives, where the pressure to perform in an unfamiliar educational system created a distinct set of needs. AI for academic challenges is another use case we learned about in our interviews, consistent with survey findings. Examples of AI applications for academic support include improving teamwork and communication skills, refining resumes, writing cover letters, and using AI to better understand classroom lectures:
\begin{quote}
    ``\textit{When I was doing teamwork, [and] communicate with somebody online, I will firstly, tell the AI what kind of thing that I want to say for my own sentence, and send it to AI, and let the AI to correct me, let the AI tell me if I have any grammar mistakes.}'' (P14)
\end{quote}

While academic and socio-cultural challenges were the most actively managed through AI, participants also touched on a more personal dimension of their experience, specifically the emotional toll of living far from home. Our participants discussed using AI for psychological matters, such as random conversations to help with loneliness in low-stakes situations:
 ``\textit{I feel a little lonely at this point. Using AI helped me to chat about and have some random conversation with AI, so it really did help me to feel less lonely at this point.}'' (P3).
 
However, a distinct psychological boundary persists among participants related to a lack of emotional nuance in AI, technical and institutional concerns, and the irreplaceable value of human empathy. In addition, we observed an internal self-judgment for using AI for emotional matters: ``\textit{I would just be creeped out. Who in their right mind had the idea to manipulate my emotions to become my friend using an AI?}'' (P12).
Findings show that there is not a significant interest in using AI in psychological challenges, especially as a long-term support. Therefore, AI is not yet considered a tool for in-depth mental health support. This hesitation is elaborated in Theme 3.

Alongside these emotional challenges, the practical demands of building a new life in the US introduced a fourth domain of AI use: logistical navigation. Participants reflected on using AI to overcome logistical challenges such as finding jobs or internships in the US, preparing for and learning about common norms in US job interviews, understanding car purchase procedures, and navigating commutes in a new country with a new transportation system. For example, P8, particularly emphasized the use of AI for what they ``\textit{want to build as a career}'' and expressed their interest in using AI for more of logistical support.

\subsubsection{Pre-arrival challenge navigation}
However, the use of AI as a support tool is not only limited to inside the US but also extends to pre-arrival processes. For example, these challenges include using AI to assist with sending emails to foreign university staff or professors, learning about immigration rules, understanding and completing foreign university applications due to language barriers, and preparing for visa appointments and document submission. We found that AI usage before coming to the US helps mitigate the stress of such tasks and stay focused on the right path: 
\begin{quote}
    ``\textit{Even before coming to the US, I started using AI, trying to get help with understanding all the process to apply for the PhD, all the documentation, to send emails... because, of course,  English is not my first language. So sometimes I'm not really sure about a lot of the information.}'' (P4)
\end{quote}
AI use before arriving in the US helped mitigate the stress of unfamiliar processes, and this extended beyond applications. P4 noted that the visa process carried the same burden: ``\textit{the visa process is just another process that requires a lot of documents. I wasn't sure what to do or what not to do.}'' Insights about using AI before coming to the US highlighted the fragmented support system that should not only help students in the US but also better support them with the procedures before arrival.

\subsection{Theme 2: External Challenges of a New Life Draw International Students to AI}

\subsubsection{Role overload and self-reliance shock encourage using AI for speed enhancement}
Participants identified role overload, due to the overwhelming number of responsibilities they must carry entirely on their own as foreigners in the US, as one of the primary drivers of their AI use.  We encouraged our participants to reflect on this overwhelming life and how it might impact their experiences as international students in the US, and the use of AI. Based on participants' experiences, we learned that students in the US face a sudden responsibility overload. This overload is due to many external factors imposed on international students. P5 compared the ecosystem in the US to their home country and highlighted that the sudden burden of life chores abroad (previously managed by family members in the home country) creates an imbalance with high-stakes tasks like finding a job or completing coursework: 
\begin{quote}
    ``\textit{In India, my mom used to cook for me, clean the dishes, do the laundry, and generally used to take care of me. I got habituated to that. So, from that culture, I came here. And here, I am cooking, cleaning, and doing all sorts of chores. With everything, it becomes a little difficult to find time to do things like this. Also, there is a pressure of finding a job, or completing a coursework assignment, and lots of things.}'' (P5)
\end{quote} 
%
In these situations, conversational AI helps mitigate this void by supporting tasks related to academic studies and logistical matters. P8 summarized this compounding pressure: ``\textit{all those things together just want you to use AI.}'' International students are in a ``\textit{constant learning process}'' (P13) that exposes them to new experiences. These new experiences require constant navigation. P13 believed that AI  provides automation that helps reduce the workload associated with overwhelming responsibilities.

All of these factors increase pressure on participants in time-limited situations; therefore, it is important for them to have access to support systems that help them with their challenges in real time:
``\textit{It's the fastest. I can get answers in 5 to 10 seconds. That's why I picked AI as my first source.}’’ (P12). 
Students are not just being impatient; they are trying to be efficient because their new life in the US has left them with no buffer time. Participants emphasized that the primary value of AI lies in its ability to mitigate time constraints arising from their newfound responsibilities. 
In this situation, the accuracy of the next step for any ongoing challenge is crucial to avoid the risks associated with trial and error, as P4 mentioned: ``\textit{I would like it to be accurate in what the next step is... not feel like I'm losing my time.}’’

In addition, participants emphasized the value of conversational AI tools such as ChatGPT, Gemini, etc., providing targeted or question-specific outputs; it helps with saving time instead of going through different resources on traditional search engines, and avoids ``\textit{wasting a lot of time scrolling}'' (P6) and manually aggregating data from ``\textit{different articles resources}'' (P10).

P9 elaborated that AI responses are sufficiently precise that users can customize their AI to provide more concrete, to-the-point responses. Such interaction can help determine the next steps more quickly. 

Among these conversations about using AI to address speed and life pressure in the US, participants reflected on the importance of the history and context stored in their AI chatbots, which helped save time and prompted AI to answer their questions. Some participants noted that, due to a long history stored in their current AI chatbot, they refrain from switching to a new chatbot because they have to provide the same context all over again: ``\textit{I keep using ChatGPT because it has my history}'' (P7).

These conversations gave us insight into how important it is for our participants to provide the AI with appropriate context to receive responses aligned with their unique life circumstances as international students.

\subsubsection{Human-human interaction inconveniences encourage using AI as a safe space}
Participants expressed a sense of being judged by people around them, fearing being thought of as a foreigner who struggles to understand the native language and constantly asks questions: 
\begin{quote}
    ``\textit{[Sometimes AI is the right choice] because AI will be more patient and will not get annoyed if you cannot understand some points that it makes. But if I'm asking questions to a real human, I am maybe afraid if she or he gets annoyed [or] if I cannot understand lots of things. That's the main reason.}'' (P7)
\end{quote}

The notion of patient AI vs. annoyed human suggests that the fear of bothering people makes conversational AI a preferred choice, helping them avoid \textit{creeping} people out and being judged. P13 highlighted that by using AI, they can find the answer or solution to their challenge
``\textit{without having to talk to people about it, or be a creep and observe how people are doing stuff.}'' (P13)

Such social distress further encourages AI use, given AI's detachment from human emotion or characteristics. Unlike human interactions, which come with the risk of social judgment and accidental or intentional oversharing data with mutual individuals in a shared social circle, the AI was perceived as a safe way: ``\textit{So I feel like it's easier if it's, like, a tool or a computer that doesn't have that attachment, doesn't know your family, doesn't know your friends, or will tell anybody else.}'' (P13)

 Another reason to use conversational AI is that friends or family, especially those in the home country, are not always readily available. Such unavailability is even more pronounced given the time zone differences between the US and participants' home countries. P13 described a situation where she needed cooking advice, but turning to AI simply because their parents were asleep in another time zone: ``\textit{my parents are in a whole other time zone, and they're probably asleep, so I don't know if this fish is good or not — let me kind of ask GPT}''.
 As P11 further emphasized the benefit of AI in these situations to \textit{`` get an immediate reply within seconds}'' (P11).
 
While a lack of easy access to family support persists (e.g., for emotional support and daily life difficulties), participants highlighted that their families often lack the localized knowledge required to navigate specific US systems.
For example, a family's knowledge of their home country's tax procedures does not directly translate to the US tax system. Consequently, international students tend to turn to conversational AI as an alternative for real-time problem solving.

\subsection{Theme 3: The Multi-Aspect Limitations in AI Ecosystem Discourage Long-Term Use}

In this theme, we elaborate on the narrative that the technical, organizational, and interactional limitations of conversational AI draw international students toward short-term use of AI. 
While participants expressed an interest in long-term guidance, the current limitations of AI shift usage towards more question-and-answer interactions. We encountered mentions of conversational AI reduced to a ``more powerful search engine’’ (P6), rather than a persistent, continuing mentorship. Participants engage with AI as an immediate remedy for solution finding: ``\textit{for, like, 99\% of the cases, it's a short-term thing. I would use it for, like, 10 minutes, and that's pretty much it.}’’ (P12).
The following sections explore deeper into the rationale for using AI for short-term tasks. 

\subsubsection{High-level, organizational limitations that discourage long-term use}

Findings highlighted that conversational AI carries high-level organizational limitations that impact how these tools are being used by international students:
``\textit{For sensitive situations, something like immigration or medical, you just have to be very cautious in those situations}'' (P8). According to P8, international students may be at risk when engaging with AI due to concerns about legal and institutional security, as they perceive their immigration status to be under constant scrutiny.
This concern extended beyond general caution. P13 described a fear tied directly to legal status: ``\textit{sharing a lot of your information is like, oh, will ICE get it? Will they come and try to take you back home?}'' For some participants, data privacy was not an abstract technical concern but a question bound up with their continued presence in the country. 
As a result, their engagement with AI becomes more short-term than long-term, a strategy to minimize their digital footprint and avoid potential long-term data-driven vulnerabilities.

For example, privacy concerns were related to the disclosure of their data to higher government authorities or to academic higher-ups, such as professors or college staff.
For our participants, even the possibility of such a breach causes significant stress. 

Beyond legal privacy concerns in the host country, participants expressed social anxiety that extends to their home-country social circle. P13, from India, highlighted the fear of data breach to their social network back home, and being misinterpreted within their original cultural context. 
P13 elaborated that they avoid long-term or in-depth engagement with AI to prevent the creation of a digital record that might contradict home country expectations or cultural norms. In addition, the state of not fully knowing how and where their data might be shared contributes to this anxiety: ``\textit{Will people get all of this information, or can my own family [in India] access it?}'' (P13)

Participants tend to be extra cautious when using AI and avoid sharing sensitive information through long conversations that require personal data, such as mental health matters or psychological problems. Therefore, participants are drawn to use AI for challenges that can be addressed through short conversations, rather than in-depth, long-term, ongoing interactions.

\subsubsection{Interaction-level limitations discourage long-term use}

While students recognize AI’s benefit for problem-solving, they find it limited in navigating the emotional challenges.
Findings highlight limitations related to AI being unfamiliar with regular human interactions, incapable of human-like emotions, human expressions, or giving out cliché answers:
\begin{quote}
    ``\textit{When you talk to a real human being, she can give you an expression, like the things that you can really feel in person. That's not what AI can do. When I go to AI, I am searching for an actual solution that I can truly apply.}’’ (P7)
\end{quote}
These limitations discourage our participants from initiating long-term, continuous conversations. This is because they believe that AI's lack of human interaction nuances hinders the conversation from progressing beyond a quick, solution-based interaction, especially when involving in-depth emotional nuances.

Following up on AI’s lack of social nuance, participants highlighted a situation in which the AI brings up social or cultural trauma.
Although prompting in native languages is sometimes used to elicit more culturally relevant responses, we encountered conflicting views on whether language prompting positively affects the social nuances of conversations. For example, P7 believed that AI expresses emotions more accurately when interacting in its native language, Chinese. P7 illustrated this with a specific example, describing how DeepSeek's responses felt ``\textit{condescending, patronizing, mansplaining,}'' evoking cultural dynamics from their home country rather than offering neutral support.

Other aspects of interaction shortcomings were related to the structure of conversations and chat tabs. P5 described a sophisticated plan to use ChatGPT as a scaffolded communication coach, to be implemented gradually. However, this plan failed due to the project structures in the AI platform: 
\begin{quote}
    ``\textit{There was a time when I planned to use ChatGPT for teaching me communication, like my frequent grammar errors and everything. So I thought, like, I would gradually go that way. But it did not work out for me to stick to the long-term plan, because maybe I was too distracted with all the other projects there.}’’ (P5)
\end{quote}

While participants initially approached AI with ambitious, long-term learning objectives, these idealized plans often failed to ``stick to the long-term initiative’’ because the AI conversation organization is too distracting with all the other projects:
``\textit{just to get ideas, and to fix grammar, and also, like, to structure my sentences better.}’’ (P12)
This suggests that long-term use is frequently displaced by short-term task completion or even abandoned. For international students, AI often shifts from being a proactive teacher to a reactive tool under the weight of a heavy cognitive load.

\subsubsection{Technical-level limitation discourages long-term use}

Participants perceived AI as failing to provide accurate outputs in certain situations. Some participants highlighted that AI sometimes sounds like a people-pleaser: ``\textit{I also think of AI, like, specifically ChatGPT, is such a people-pleaser tool}’’ (P4).

AI’s responses are sometimes perceived as overly biased based on the context the user provides; such a lack of honesty in AI responses is perceived as a discouraging factor in continuing to use AI for ongoing conversations over longer periods. 

In addition, some participants expressed annoyance during interactions with AI that enforce critical outputs. P5 described a situation in which she asked AI for feedback on a conversation with her advisor. In turn, the AI's criticism felt overly abstract and unrealistic. Such forced criticisms makes P5 to ``overthink about this stuff'' (P5)
Similarly, participants expressed frustration that AI sometimes produces random responses or hallucinations. This issue was highlighted by algorithmic limitations in which AI loses context of the conversation as the conversation grows too long. Some participants also expressed frustration that AI sometimes fails to follow instructions given the provided context, or, in some cases, mixes previous chat context with the current conversation. P13 captured this frustration directly, noting that managing hallucinations required constant correction:
\begin{quote}
    ``\textit{I need to keep going back, like, don't hallucinate data, ask me if you need anything. I feel like at that point, I'm talking to it like a child.}'' (P13)
\end{quote}
In these situations, participants either switch to another AI tool, open a new chat, or manually prompt the model to stay on track. This activity suggests that technical issues in AI models also contribute to the avoidance of AI for long-term tasks.

\subsection{Summary of Interview Findings}
The interview findings provide nuanced explanations for patterns observed in the survey, identifying the specific gaps due to life differences that drive AI adoption among international students. While the survey highlighted a preference for quick, task-oriented support, the interviews further clarify that this is partly a response to overwhelming life responsibilities and a state of high self-reliance. Thus, participants are drawn to use AI to manage a sudden increase in responsibilities and handle the time-consuming process of manual information seeking. 
AI is often considered a patient, non-judgmental friend to avoid the social cost of bothering others, such as advisors, peers, and native speakers, with repetitive questions. 

However, there is a hesitation to use AI for emotional or psychological support. This is not merely a preference for human input, but a response to perceived institutional and interactional risks that play a great role in our participants' lives. These include concerns about data privacy, especially when data affects visa/immigration status, and the AI’s lack of social nuance.

Considering the various reflections from our participants, our interview findings suggest that while AI is primarily used as a first-aid tool for immediate challenges, there is potential and a desire for proactive, long-term AI-powered support tailored to international students' unique challenges. However, it is currently hindered by AI's organizational, interactional, and technical shortcomings. These gaps directly inform the design considerations discussed in the following section.

\section{Design Ideas and Considerations}\label{design-ideas}

Following up on our participants' reflections and their interest in a long-term AI support tool, we encouraged them to suggest how this desire should be achieved. In the following sections, we present suggestions and key considerations for creating an AI-powered support tool, particularly tailored to the needs of international students (IRQ3).

\subsection{Higher-level institutional considerations}

\begin{itemize}
    \item \textbf{Privacy Disclosures and Transparency as Trust-Builder} -- Provide real-time disclosures where the AI informs of user data sensitivity, explains how data is handled, and the reasoning behind it.
\end{itemize}
Given their major concerns about data privacy and trust, our participants strongly recommended that any AI-powered tool tailored to the needs of international students should implement robust privacy procedures. 
Among these considerations, the disclosure of privacy protocols has been highlighted. Recommendations for privacy include transparency on how data is handled, who it is shared with, what policies are followed, and how users have agency over their data privacy.

These suggestions underscore the importance of users' awareness of AI privacy policies and how they function. P8 highlighted suggestions on how AI can play a part in enabling this awareness using warnings or reminders during conversations in real time:
\begin{quote}
    ``\textit{[Using] warnings, or kind of an information, or a pop-up [...] while you're typing out the prompt, they (AI) should just, go through the prompt and assess it, and then give you this particular warning where, you know... just be careful if this might be used for training, if this is just sensitive data just giving you the reminder where... you know, before sharing a sensitive information, they should probably provide, some sort of warning.}'' (P8)
\end{quote}
This awareness should also include awareness of one's control over how the shared data is used. Participants mentioned they take caution for themselves and apply privacy boundaries for their personal data while interacting with AI, by being selective in sharing types of data; however, participants suggested that it should also be a responsibility of AI to inform the users on what data should or shouldn’t be shared to help its users protect their privacy and mitigate their fear of data breach.

In addition, participants highlighted the reliability of sources, including a desire for transparency regarding the sources used in AI outputs. Participants often cross-reference AI outputs against trusted sources and seek clear attribution to verify the information's reliability. Another big concern in this area was safeguards for user data, including a desire for direct control and manual ``check-off'' features to manage their data security, such as using their input to train the AI. Ultimately, participants expressed that these disclosures or any features should be ``easily visible'' and accessible to users.  

\begin{itemize}
    \item \textbf{Integration in Already existing University Support Systems} -- Use local machine/university server to enable a safe space and overcome the privacy and trust issue
\end{itemize}
Participants suggested using local machines for the AI-powered tool. One suggestion in this regard is to integrate the AI tool into the existing support systems provided by the universities, such as the International Students Office (ISO) or International Students Services (ISS), Academic Advising \& Faculty Office Hours, and Career Centers; doing so will enable it to run on the university's local machine. 
As P9 explained, running AI on a local machine ensures that when a user asks a question, ``\textit{it does not go to OpenAI servers and then to train the servers. It's basically on their own machine... I can provide it a bunch of context over time, because I know that it's not going to use my data for anything.}'' 

By hosting such a system internally (e.g., ISO/ISO), universities can provide institutional assurance of the safety of students' privacy as P1 noted: ``\textit{if I see that many people that have a hand in developing those tools can reassure me on safety of my data, I will be more inclined to use them.}''. 


One of the deal-breakers participants mentioned in this tool is the cost, given international students' financial hardships.  Integration into university support systems can provide access to a free support tool tailored to international students.

\begin{itemize}
    \item \textbf{Human-AI Combination Workflow} -- Create a human-in-the-loop system, for when queries become high-stakes.
\end{itemize}
This integration should enable a support environment that leverages both AI and human resources within the university system. As participants highlighted, the AI component will complement the current support system, and vice versa. P9 explained that ``\textit{if you’re not satisfied, you can push for human support. That way, then you can maybe schedule a time.}''
P4 further elaborated with an example to illustrate how a new system, combining a human and AI component, is needed to assist with task overloads:
\begin{quote}
    ``\textit{A new system where you can integrate both people working at the international office, but also AI tracking your progress on the PhD in the institution that you are... On different aspects, like your academic aspect, your financial aspect, the taxes, and your [immigration] status.}'' (P4)
\end{quote}

Participants endorsed the integration of such AI tools into already existing university support systems not only because it helps with the privacy and trust issue, but also because it could help reduce the fragmentation issue in support systems, rather than requiring students to diagnose their needs and seek out disparate resources independently. P4 noted that an integrated system would offer a centralized point of access and use existing student data to contextualize support: ``\textit{integrate what it knows about me and my current challenges}'' (P4), ensuring that for high-stakes issues, students can still reach a reliable human authority ``\textit{to find a specialist} (P1).


\begin{itemize}
    \item \textbf{Community Building} -- Enable International Students to find and connect to their relevant communities.
\end{itemize}
The AI tool should enable community engagement by incorporating a hybrid human-AI combination. P4 emphasized the importance of community building for international students, to be able to find the people they could sympathize with and ask questions or get proper, relevant answers, since finding such good communities can be challenging: ``But also, having a community there. Like, not only staff, not only AI, but also, like, you as an individual being there.’’ (P4)

\subsection{Multi-Dimensional Customization}
Beyond the specific considerations for privacy assurance, our participants highlighted the importance of AI tool customization, which could enable them to personalize their support based on their context and needs, background, and language proficiency. For instance, P1 noted a significant gap in current tools regarding linguistic customization, stating, ``\textit{when I chat with it in my native language, Farsi, it’s not that intelligent as in English... I wish it would have been more reliable in my first language.}'' Beyond language, participants emphasized the need for the tool to maintain a history of their unique circumstances to provide relevant support. As P4 described, an ideal assistant should  ``\textit{integrate what it knows about me and my current challenges}'' into its responses.

This customization should involve a system that can solve the problem without having students manually input their context every time. This level of personalization would allow the tool to act as a proactive guide, or as P7 put it, ``\textit{like a friend that's always there and always able to help you out}''

Suggested customizations should consider the following aspects:

\begin{itemize}
    \item \textbf{Domain-Specific Modalities} -- Instead of one domain focus, suggest specialized domains that users can toggle between.
\end{itemize}
The AI tool should cover all four domains of challenges, allowing international students to customize which domains their AI focuses on. P5 highlighted that this customization should be multi-aspect, where the user should be able to have more than one domain-customized conversation tab:
\begin{quote}
    ``\textit{If I feel like I want some, you know, mental support at this point of time from a psychological domain. But maybe, like, in 2 or 3 days, I want a help with social-cultural challenge, where I want to describe what I'm doing and all. So I should always be able to switch between them.}’’ (P5)
\end{quote}
Such customization can take a more automated, suggestive form, in which an AI tool would prompt users to select the domains they want to focus on to initiate conversations.

\begin{itemize}
    \item \textbf{Socio-Cultural Context Grounding} -- The AI tool should collect social/cultural context from users and ground responses in them.
\end{itemize}
The AI tool should understand its users' context to move beyond generic outputs and provide culturally and socially aware responses. For instance, P9 highlighted the AI's usage depends on its ability to relate based on where the student is coming from:
``\textit{You being an international student, it should know where you're coming from, so that it can maybe relate based off of [that], because I'm sure all countries are different, like, the way we are placed in Nigeria is completely different from the way you're operating in Iran}’’ (P9)
This is especially important when navigating cross-cultural differences, as the AI must account for the fact that a student's home country (e.g., Nigeria) may operate under entirely different standards than inside the US.

\begin{itemize}
    \item \textbf{AI Role Customization} -- Enable the AI to be able to be customized and perform multiple roles.
\end{itemize}
The role of the AI tool should be customizable. For example, participants described using AI as a friend or assistant, particularly for long-term use. A friend or companion for loneliness was mentioned by P3, where AI can help with their loneliness, through random conversations: ``\textit{Using AI helped me to, like, to chat about and have some random conversation with AI, so it really did help me to feel less lonely at this point.}’’ (P3)


Building on the roles of AI tools, we found suggestions related to AI playing the role of a coach for long-term use: 
\begin{quote}
    ``\textit{I think it could definitely be a coach...Because it has the knowledge base, giving us specific instructions, and basically coaching us on, like, the route to take to finish the project, you know?}'' (P12)
\end{quote}
This coaching role should extend beyond academic into soft skill development and professional navigation. P2 highlighted that they use AI for communication skill enhancement, through ``\textit{practice problems}'' while P13 tends to use AI as a coach to offer advice: ``\textit{Like, how do I approach speaking to this professor.}'' (P13)

Other roles highlighted were a \textit{tutor}, who provides a learning environment; and a patient tutor for practicing English communication with the user, where no judgment is made about the student's current language proficiency. This is particularly important for mitigating the cultural shock inherent in the international student experience: ``\textit{due to language, or due to cultural shift or change, how to adjust because of that. So, the tool will definitely help with that cultural shock and communication challenges.}'' (P9), 
or an educator on logistical and bureaucracy topics: ``\textit{if there was a dedicated chatbot which went into, the nuances of, for example, filling out certain forms}'' (P10).

When asked to elaborate further on their suggestions for the tutoring role, participants described how such tutoring could be implemented in the AI tool. P5 suggested the AI tool can use time-sensitive micro-lessons or chat-based educational sessions. The reason for using micro lessons is to accommodate the overwhelming and busy lives of international students:
\begin{quote}
    ``\textit{Here, I am cooking, cleaning, and doing all sorts of chores. With everything, it becomes a little difficult to find time to do things like this, where I'm learning something, or sticking to a course. Also, there is a pressure of finding a job, or completing a coursework assignment, and lot of things. So, for me personally, if the app is giving me smaller time period of effective lessons, that would be really great, because then I'll not feel like, oh, this will take this much of time, and I'll not push it to tomorrow}'' (P5)
\end{quote}
The emphasis on time-sensitive micro-lessons was also inspired by current educational apps, such as Duolingo, which some participants used. 

\subsection{Interaction Paradigms: From Reactive to Proactive}

Participants emphasized that for AI to truly integrate into the high-pressure lifestyle of international students, it needs to move beyond a reactive website and become a mobile-first, proactive companion. Throughout conversations, participants often referred to the AI tool as an app that users can easily access, not only on a website, but also on their mobile phones. 

\begin{itemize}
    \item \textbf{Proactive Life Cycle Responsibility Mentorship} -- The AI should transition from a passive informative tool to an active companion that knows and anticipates its user's needs based on the international student's context.
\end{itemize}
P8 described the need for this shift as a system that acts as a profile that already knows a user's progress, checks on them, and helps them stay on track with their responsibilities as international students: ``\textit{Like, hey, you need to do your taxes. Here’s who you can talk to. Like, something more proactive.}'' 

This mobile-first approach is particularly crucial for international students who face a high number of challenges. As P5 suggested, the app should be a ``\textit{constant companion}'' that initiates help. For example, ``\textit{As we enter into the app, as an international student, the app should ask me which domain I want to focus on.}''

Instead of waiting for a prompt, participants envisioned a system that keeps users \textit{on track} through interaction delivery modalities such as automated reminders and pop-up notifications about critical timelines—including tax filings, immigration deadlines, academic progress, and bureaucratic document submissions. To sustain long-term engagement and mitigate the cognitive load of these tasks, participants also suggested the incorporation of gamification features: ``\textit{a little bit of gamification, like, right now on Duolingo}'' (P5) to encourage consistent engagement through check-ins and motivational encouragement.

\begin{itemize}
    \item \textbf{Restructuring of the Conversational Architecture} -- The AI should provide information in chunks and wait for user validation before proceeding.
\end{itemize}
Beyond the delivery method, participants identified a need for a fundamental restructuring of the conversational architecture. A recurring critique of current models was the `information dump,' long, monolithic responses that often result in hallucinations or irrelevant biases. Instead, students preferred incremental outputs delivered step-by-step, with AI actively asking follow-up questions. The rationale for an incremental, iterative approach was to mitigate hallucinations and biases by enabling the AI to extract more relevant social, cultural, professional, and other contextual information from the user. As a result, it would be more accurate to provide a more manageable flow of information.

\section{Discussion}

Our paper's contribution is survey and interview findings that illustrate how and why international students in the US tend to use conversational AI to address their unique challenges. These findings contribute requirements and suggestions for creating and improving AI-powered support tailored to the needs of international students. 

\subsection{Place of Conversational AI in the International Students Life Cycle}

In line with previous studies~\cite{oduwaye2023trend,bethel2020cross}, both survey and interview data show that international students face four main challenge domains: academic, socio-cultural, logistical, and psychological. These challenges come in the form of daily life demands and academic pressure for an international student, leading them to rely on AI tools, due to the tools' easy access~\cite{cena2026studying}.
While universities provide official support through communities, or traditional university services (advisors, counselors), our findings suggest significant limitations in these traditional services.
Therefore, international students adopt AI as a \textit{first-aid} tool using its 24/7 availability to compensate for the voids in a traditionally fragmented support ecosystem~\cite{bethel2020cross,zhang2018chinese,cogan2024taboo}. 
We confirm this fragmentation in our paper; however, we extend this discourse by highlighting creative ways in which AI helps mitigate students' challenges in the high-stress demands of life in a foreign country. 

In particular, participants often experience cognitive fatigue due to the shock of self-reliance when living alone and constantly learning new things (P5, P13). In this state, the manual search for a support resource, scrolling in traditional search engines, or looking for a support community is not just a time-waster; it is a mental burden they find difficult to afford. Participants' suggestions to address this issue by incorporating micro-lessons in an AI-powered tool are supported by the ``Principle of Least Efforts'' discussed by Case and Given~\cite{case2016looking}. According to this principle, people tend to choose the path that requires the least amount of effort and prioritize immediate access over exhaustive traditional sources to achieve a satisfactory outcome.
Breaking complex information into small units is beneficial to prevent becoming overwhelmed during periods of high stress~\cite{sweller2020cognitive}, such as adapting to the new daily life changes and academic system.

Moreover, the suggestions related to proactive AI that anticipates users' needs highlight the importance of minimizing the mental overload that individuals experience, by reducing the need to initiate action~\cite{pejovic2014anticipatory,pejovic2015anticipatory}. Especially in the context of our paper, doing so is invaluable as international students experience immense difficulties with language barriers, social isolation, academic challenges, logistical tasks, etc. 

\subsection{Reflecting on Concerns Regarding AI Usage}

The interviews clarify the psychological boundary identified in the survey: students' hesitation to use AI for emotional support is not merely a preference for humans, but a response to perceived institutional and interactional risks. These include concerns over data privacy (particularly regarding visa status), fear of cultural hallucinations, and the AI’s lack of social nuance, also supported by previous work~\cite{wang2023exploring}. Our paper, however, highlights these concerns from a more nuanced lens, contextualizing these risks within the unique lived experiences of the international students.

Recent work~\cite{cogan2024taboo,lee2025artificial} has studied the hesitation to use AI for mental health, indicating a restrictive boundary set by people regarding the sharing of in-depth or personal data with AI. For our participants, privacy concerns appear to be a significant one. International students are not only concerned about data breaches or privacy; they are in a state of vulnerability, where a single misstep (e.g., legal, academic, or social) could lead to the loss of their visa or legal status, as indicated by previous work~\cite{acar2025welcome}. Therefore, this makes the privacy issue in AI much more than just a tech concern; it’s a survival concern.

These states of vulnerability among international students partly explain the psychological boundary identified in our findings: students are hesitant to share in-depth personal data not only because AI lacks human emotion understanding, but also because of the potential misuse of their digital footprint. 
Particularly, international students' fear about data privacy marks a critical indicator of contextual stakes involved in designing AI-powered support tool for this demographic. In HCI research, the importance of understanding the nuanced life experiences of users has been extensively emphasized, particularly when designing for marginalized populations\footnote{International students are recognized potentially as a marginalized group~\cite{tavares2024feeling}.} who face unique systemic challenges~\cite{lee2025into,gautam2025towards,oluwatuyi2026collectively,hassan2026asl}.

Moreover, the risk of cultural hallucinations/misinformation and encoded biases~\cite{bender2021dangers} further marginalizes these users, in which case, the harms of AI bias fall disproportionately on those already navigating a foreign system.
If the AI does not understand their cultural context and background correctly, it means not just a hallucination, but a reinforcement of their marginalized status within a support system.
The interview findings show that cultural bias is not only a matter of inaccurate information, but also a matter of how AI interaction is felt by students navigating between cultural contexts. Participants described moments where AI responses seemed to reproduce inconvenient social dynamics from their home cultures (e.g., stereotypes) rather than offering neutral support. 

There has been a large discourse on LLMs culturally and socially biased~\cite{liu2024understanding,lee2025into,fortunati2022people}, particularly enforcing Western notions in their outputs~\cite{agarwal2025ai}. For international students, such biases especially matter because the same AI response can be interpreted through multiple cultural frames at once: the norms of the student's home country, the norms of the host country, and the assumed norms embedded in the AI system~\cite{wang2023exploring,ma2025exploring}.

 Prior work on LLMs has warned that training data can encode dominant cultural assumptions and reproduce representational harms \cite{bender2021dangers,agarwal2025ai}. HCI research similarly emphasizes that systems designed around Global North assumptions may fail to reflect the values, preferences, and lived realities of users from non-Western or migrant backgrounds \cite{nourian2025invest,nourian2025cultural}. In our study, this concern appeared when participants wanted AI to understand where they were coming from, but also feared that AI responses could be inaccurate, stereotyped, or socially triggering. Therefore, cultural grounding in AI support tools should not simply mean adding country labels or translating responses. It requires careful attention to how advice is framed, what norms are assumed, and how solutions are framed and conveyed across socio-cultural contexts.

%


\subsection{Reflecting on AI Usage as The Social Buffer and the Vulnerability Trade-off}

International students use AI to address their unique challenges and are generally satisfied with it, despite its limitations~\cite{wang2023exploring}.
Our paper showed a social safety lens not captured in the survey: AI is perceived as a patient, non-judgmental buffer to avoid the social cost of bothering peers, native speakers, or advisors with repetitive questions. 
AI output is not usually treated as the final answer to participants' queries, but as a mediator to lower the stress associated with their challenges. For example, for international students who are non-native English speakers, findings suggest that AI functioned as a language support tool, helping refine grammar and expression in ways that may reduce language-related awkwardness. Similar indications have been highlighted in the work of Cena et al.~\cite{cena2026studying}, which shows that AI is viewed as a tool that reduces language disadvantage by helping students express what they know more effectively.

Although our survey showed a strong preference for humans, interviews highlighted a ``fear of being a creep'' (P13) or ``bothering'' people (P7, P3).
We discuss that students are not using AI because they trust it perfectly or that AI does a great job at all times; they are using it to manage different types of risk such as being misjudged as an irritating person. This usage pattern aligns with previous research highlighting that AI does provide a ``safe space'' to express thoughts and seek support~\cite{chandra2025lived}.
While students express a need for long-term AI integration, they tend to trade off the social risk of human judgment against the technical risk of AI inaccuracy.

In this case, AI acts as a social buffer that, in the short term, students use to rehearse or check information to avoid appearing foolish in front of others. AI is not replacing humans but preparing students for humans. It is a low-stakes practice opportunity for a high-stakes social reality.

\subsection{Reflecting on Intent for Hybrid AI Support \& Sense of Community}

Past studies have emphasized the notion of human-AI collaborations in various domains such as healthcare~\cite{albikawi2025nursing}, marginalized groups education~\cite{hassan2026asl}, and migration~\cite{lee2025into}, emphasizing the inclusion of end-users in the research, design, and utility cycle. 
This inclusion ensures that AI outputs align with the mental models of their target users; especially, in the context of migrants, there is often a fundamental misalignment in how migrants and institutional support systems conceptualize the adaptation process~\cite{Truong2024enhancing}
Similarly, our paper incorporated the insights and lived experiences of international students, as our end-users, ensuring an inclusive and comprehensive understanding that aligns with their needs and expectations.

AI is not seen as the only stand-alone way to address issues, but a complementary tool that can help bring together the fragmented pieces of support~\cite{chandra2025lived}. 
Human-AI combination was emphasized by interview participants following the importance of human support, enabling ``a sense of community'' or ``sense of belonging''. In general, for immigrants, relevant communities function as communication tools, a critical way that expands and sustains their social networks, mitigates the challenges of being a foreigner, and prepares for the early settlement in the host country~\cite{dekker2014social}. As for international students, in addition to the aforementioned points, they use online communities to ease emotional challenges such as loneliness and cultural stress, seek academic and everyday information, and connect with peers who share similar experiences, thereby fostering a sense of belonging~\cite{baines2022social,dekker2014social}.

Given the importance of community support and connection for international students, we discuss that the AI-powered tool will benefit by serving as a mediator to help find and build relevant communities. Especially doing so plays a crucial part, as it helps with the notion of human-in-the-loop, suggested by our interview participants. In this scenario, AI does not replace humans' lived experiences but instead enables a hybrid support environment and acts as a triage agent, handling lower-stakes questions while helping users seek human-led communities to learn from lived experiences (e.g., other international students, student organizations, or advisors).

\subsection{Limitation and Future Work}

In the interview study, we wanted our participants to reflect on their AI use based on their prior experience, to avoid limiting them to any specific task or scenario. We benefited from the valuable insights of the students, gathering ideas and recommendations for improved AI-powered support tailored to their unique needs. However, since we did not ask participants to use AI chatbots on the spot or to complete specific tasks, some of our participants might have forgotten to mention some aspects of their AI usage experience. This limitation is common in interview studies. Thus, Future research can benefit from building on our findings using task-oriented user studies for more narrowed-down, in-depth insights.
Moreover, future work can continue this line of research by conducting group study sessions, such as workshops and focus groups, and also consider longitudinal procedures to capture detailed data on AI usage and feedback by international students.

In addition, while valuable, our work only shares the perspectives of students. There is an opportunity to learn more by conducting research with other stakeholders or entities responsible for supporting international students, such as International Student Offices at universities.

Finally, to reduce any bias in our data interpretation (e.g., cultural bias, professional background bias, etc.), the researchers carefully conducted a data-centric analysis, only grounding interpretations in participants' discourses rather than personal assumptions. We verified findings using comparison with literature~\cite{firestone1982approaches} and holding discussions among the team to collaboratively review the data themes. This tactic helped to identify potential misinterpretation of data.
Future work could improve these research processes by involving more international student researchers and cross-cultural HCI and AI experts, both in data collection and analysis, enhancing researchers' contextual competence through immersion.

\section{Conclusion}

We conducted a mixed-method study using a survey (n=60) and semi-structured interviews (n=14) with international students in the US to explore their adoption patterns and perceptions of conversational AI for navigating challenges of cross-cultural adaptation. 
Our work contributes a nuanced understanding of how international students use AI to mitigate the existing limitations in a fragmented institutional support ecosystem.

We found that while AI has become a ubiquitous digital first-aid tool for immediate academic, logistical, and socio-cultural challenges, its adoption for mental health or psychological issues (e.g., engaging with AI for emotional support) is met with more hesitation. This hesitation is driven by transnational privacy concerns and the lack of human expression that feels insufficient for deep emotional needs.

 Findings show an interest in incorporating AI into the traditional support systems and transforming it from a tool for short-term help into a long-term support companion. Such evolution requires a design shift in which AI is not a standalone replacement for human support but instead adopts a human-in-the-loop approach to prioritize local institutional integration, community mediation, and data transparency. By identifying these boundaries and requirements, our study paves the way for designing AI-powered support systems that are not only functionally efficient but also institutionally and socially aligned with the unique vulnerabilities of the international student community in the US.

\section*{Disclosure Statement}
We confirm that there are no relevant financial or non-financial competing interests to report.

\begin{acks}
This material is based upon work supported by the National Science Foundation under Award No. DGE-2125362. Any opinions, findings, and conclusions or recommendations expressed in this material are those of the author(s) and do not necessarily reflect the views of the National Science Foundation.
\end{acks}

\bibliographystyle{ACM-Reference-Format}
\bibliography{references}

\appendix
\section{Appendix: Survey Findings Tables}\label{apendix}

\begin{table}[h] 
\caption{Summary table of the types of challenges international students face and their prevalence}\label{challenges}
\begin{tabular}{|l|l|l|}
\hline
\multicolumn{1}{|c|}{\textbf{Challenge domain}}                                       & \multicolumn{1}{c|}{\textbf{Percentage (N=56)}} & \multicolumn{1}{c|}{\textbf{Brief description / example}}                                                                              \\ \hline
\begin{tabular}[c]{@{}l@{}}Socio-cultural, \\ language, \\ communication\end{tabular} & 75\%                                                & \begin{tabular}[c]{@{}l@{}}Making friends, navigating \\ US social costumes, language \\ barriers (accent, speed, idioms)\end{tabular} \\ \hline
Academic                                                                              & 43\%                                                & Understanding studies, workload,                                                                                                       \\ \hline
Psychological                                                                         & 39\%                                                & \begin{tabular}[c]{@{}l@{}}Managing stress, mental health, \\ missing family and loneliness\end{tabular}                               \\ \hline
Logistical                                                                            & 29\%                                                & \begin{tabular}[c]{@{}l@{}}Filing taxes, securing jobs, \\ housing, banking\end{tabular}                                               \\ \hline
\end{tabular}
\end{table}

\begin{table}[h]
\caption{Support types that international students prefer}
\begin{tabular}{|l|c|l|}
\hline
\textbf{Preferred Support Type}                                                 & \multicolumn{1}{l|}{\textbf{Frequency (N=58)}} & \textbf{Percentage (N=58)} \\ \hline
Experiences and advice from other international students                        & 33                                             & 56.90\%           \\ \hline
Help from more senior students, an advisor, a counselor, etc.                   & 30                                             & 51.72\%           \\ \hline
Trustworthy information, resources, and cultural facts about living in the U.S. & 28                                             & 48.28\%           \\ \hline
Easy and quick support/help/advice for practical and social issues              & 26                                             & 44.83\%                    \\ \hline
Support from digital tools or online resources             & 26                                    & 44.83\%           \\ \hline
Support for mental health or stress management                         & 24                                             & 41.38\%           \\ \hline
\end{tabular}
\end{table}

\begin{table} 
\caption{caption for this first Kruskal}\label{kruskal-first}
\begin{tabular}{|l|l|l|l|}
\hline
\textbf{Variable} & \textbf{H-Statistic} & \textbf{p-value} & \textbf{Finding} \\ \hline
\textbf{Usefulness}          & 15.14                & 0.002            & Significant      \\ \hline
\textbf{Intent of Continuation}        & 11.27                & 0.010            & Significant      \\ \hline
\textbf{Satisfaction}        & 6.74                 & 0.08             & Not Significant  \\ \hline
\end{tabular}
\end{table}

\begin{table}[]
\caption{Students' intent to continue using AI, their satisfaction rates, and perceived AI usefulness across the four challenge domains.}\label{intent-use-satis}
\begin{tabular}{|c|l|l|l|}
\hline
\textbf{Challenges}     & \textbf{Intent to Continue} & \textbf{Satisfaction}                                          & \textbf{Usefulness}                                            \\ \hline
\textbf{Socio-cultural} & Median=4, Mean=3.8          & \begin{tabular}[c]{@{}l@{}}Median=4, Mean= 3.49\end{tabular} & \begin{tabular}[c]{@{}l@{}}Median=4, Mean= 3.8\end{tabular}  \\ \hline
\textbf{Logistical}     & Median=4, Mean=4.22         & \begin{tabular}[c]{@{}l@{}}Median=4, Mean= 3.78\end{tabular} & \begin{tabular}[c]{@{}l@{}}Median=4, Mean= 4.04\end{tabular} \\ \hline
\textbf{Academic}       & Median=4, Mean=4.38         & \begin{tabular}[c]{@{}l@{}}Median=4, Mean= 3.88\end{tabular} & \begin{tabular}[c]{@{}l@{}}Median=4, Mean= 4.27\end{tabular} \\ \hline
\textbf{Psychological}  & Median=4, Mean=4            & \begin{tabular}[c]{@{}l@{}}Median=4, Mean= 3.55\end{tabular} & \begin{tabular}[c]{@{}l@{}}Median=4, Mean= 3.63\end{tabular} \\ \hline
\end{tabular}
\end{table}

\begin{table}[] 
\caption{caption for this post hoc}\label{post-hoc-useful}
\begin{tabular}{|l|c|l|}
\hline
{\color[HTML]{434343} \textbf{Cross-Domain Comparison for Usefulness}}           & \multicolumn{1}{l|}{{\color[HTML]{434343} \textbf{Bonferroni P-value}}} & {\color[HTML]{434343} \textbf{Finding}} \\ \hline
{\color[HTML]{434343} Sociocultural vs Academic}      & {\color[HTML]{434343} 0.0048}                                           & {\color[HTML]{434343} Significant}      \\ \hline
{\color[HTML]{434343} Academic vs Psychological}      & {\color[HTML]{434343} 0.0405}                                           & {\color[HTML]{434343} Significant}      \\ \hline
{\color[HTML]{434343} Sociocultural vs Logistical}    & {\color[HTML]{434343} 0.5961}                                           & {\color[HTML]{434343} Not significant}  \\ \hline
{\color[HTML]{434343} Sociocultural vs Psychological} & {\color[HTML]{434343} 1}                                                & {\color[HTML]{434343} Not significant}  \\ \hline
{\color[HTML]{434343} Logistical vs Academic}         & {\color[HTML]{434343} 0.2834}                                           & {\color[HTML]{434343} Not significant}  \\ \hline
{\color[HTML]{434343} Logistical vs Psychological}    & {\color[HTML]{434343} 0.5525}                                           & {\color[HTML]{434343} Not significant}  \\ \hline
\end{tabular}
\end{table}

\begin{table}[]
\caption{caption for this post hoc}\label{post-hoc-continue}
\begin{tabular}{|l|c|l|}
\hline
\textbf{Cross-Domain Comparison for Intent of Continuation} & \multicolumn{1}{l|}{\textbf{Bonferroni P-value}} & \textbf{Finding} \\ \hline
Sociocultural vs Academic                                   & 0.0054                                           & Significant      \\ \hline
Sociocultural vs Logistical                        & 0.1602                                           & Not significant  \\ \hline
Sociocultural vs Psychological                     & 1                                                & Not significant  \\ \hline
Logistical vs Academic                             & 1                                                & Not significant  \\ \hline
Logistical vs Psychological                        & 1                                                & Not significant  \\ \hline
Academic vs Psychological                          & 1                                                & Not significant  \\ \hline
\end{tabular}
\end{table}

\end{document}